\documentclass[10 pt, conference]{IEEEtran}

\usepackage{epsfig,graphicx,color,amsmath,amsfonts,bm,balance,tikz,array,multirow,fancyhdr,float, blindtext,algorithm,algpseudocode,url,hyperref,graphicx,subfigure}

\title{\LARGE \bf
Model Order Estimation in the Presence of multipath Interference using Residual Convolutional  Neural Networks
}

\author{Jianyuan Yu, William W. Howard,  Yue Xu and R. Michael Buehrer
\thanks{$*$This work was supported by Wireless @ Virginia Tech}
\thanks{$^{*}$}
\thanks{$^{\dagger}$todo%
       } \\
$^{}${\it Wireless @ Virginia Tech},  Bradley Department of ECE, Virginia Tech, Blacksburg, VA 24061 \\

\{jianyuan, wwhoward, xuyue24, buehrer\}@vt.edu
 
}

\begin{document}

\maketitle
\thispagestyle{empty}
\pagestyle{empty}


\begin{abstract}

Model order estimation (MOE) is often a pre-requisite for Direction of Arrival (DoA) estimation. 
Due to limits imposed by array geometry, it is typically not possible to estimate spatial parameters for an arbitrary number of sources; an estimate of the signal model is usually required. 
MOE is the process of selecting the most likely signal model from several candidates. 
While classic methods fail at MOE in the presence of coherent multipath interference, data-driven supervised learning models can solve this problem. 
Instead of the classic MLP (Multiple Layer Perceptions) or CNN (Convolutional Neural Networks) architectures, we propose the application of  Residual Convolutional Neural Networks (RCNN), with grouped symmetric kernel filters to deliver  state-of-art estimation accuracy of up to 95.2\% in the presence of coherent multipath, and a weighted loss function to eliminate underestimation error of the model order. 
We show the benefit of the approach by demonstrating its impact on an overall signal processing flow that determines the number of total signals received by the array, the  number of independent sources, and the association of each of the paths with those sources . Moreover, we show that the proposed estimator provides accurate performance over a variety of array types, can identify the overloaded scenario, and ultimately provides strong DoA estimation and signal association performance.

\end{abstract}


\section{Introduction} 

Passive DoA estimation plays a vital role in a number of localization, radar and communication systems. 
DoA estimation can enable a system to localize and track sources in both azimuth and elevation. 
Typically, DoA estimates are acquired through model-based methods \cite{krishnaveni2013beamforming} such as beamforming, matrix pencil approaches, or super resolution estimators such as Multiple Signal Classification (MUSIC) \cite{vanderveen1996joint} and Estimation of Signal Parameters via Rotational Invariant Techniques (ESPRIT) \cite{wong1997uni}. 
However, each of these techniques comes with requirements. 

Beamforming techniques and ESPRIT only consider single-source DoA estimation, while in the more general case the channel may be occupied by several signals. 
The matrix pencil technique provides limited resolution and accuracy. 
MUSIC and derivative techniques suffer when the manifold\footnote{The array manifold represents the array response for given spatial parameters.} is inaccurate, which can be caused by subtle effects such as array manufacturing imperfections, coherent multipath\footnote{Coherent mulitpath is distinguished from non-coherent multipath in that the mulitpath is temporally correlated within a symbol interval thus distorting the spatial covariance matrix of the received signal.} as well as mutual electromagnetic coupling. 
MUSIC also suffers from inaccuracy due to peak detection.
All of these discussed DoA techniques also suffer from an \emph{a priori} requirement on MOE. 

The widely used Akaike Information Criterion (AIC) and  Minimum Description Length (MDL) approaches \cite{wax1985detection} are both statistical inference tools for model selection. 
Unfortunately, both techniques degrade sharply  for small sample sizes and/or non-white spatial noise. Additionally, the presence of coherent multipath has a large negative impact on these approaches, as we will show.
It is proposed in \cite{wu1994source} to use the Gerschgorin disks method to cluster the signal and noise eigenvalues after recasting the covariance matrix into a sparse matrix with non-coherent signals.
This method requires a heuristic cluster threshold, which results in poor generalization performance across different array geometries and does not address the problem of coherent multipath. 
More recent work \cite{qiu2015method} estimates the non-coherent multipath signal model by purposefully overestimating, then finds valid peaks on the corresponding MUSIC spatial spectrum. However, as the model order increases, this technique begins to lose accuracy.

Data-driven Deep Learning (DL) models have been a promising solution to overcome the aforementioned issues like array-dependent parameters or coherent interference due to  their powerful ability to learn high level features from high-dimensional structured data at multi-scale resolutions. 
As two examples of early supervised learning techniques applied to DoA estimation, \cite{pastorino2005smart} uses SVM to  estimate DoA even when the manifold is unknown, and \cite{el1997performance} was the first to propose a one layer neural network with radial-basis functions to estimate DoA. 
More recently,  Convolution Neural Networks (CNN) were first adopted in \cite{chakrabarty2019multi} for multiple speaker DoA estimation. 
It was shown that a CNN can learn the representational features of spectral images obtained via the short-time Fourier transform (STFT), and can provide performance better than MUSIC.
Unfortunately, the computational complexity required to generate the spectral images limit the model's real-time applicability. 
On the other hand, \cite{papageorgiou2020deep} \cite{papageorgiou2020direction} \cite{papageorgiou2020fast} use the covariance matrix as the input feature with uncorrelated signals, since it is more compact and easy to generate from the received signal. 
One issue shared by all of these techniques is the required knowledge of the MOE. 

Additionally, deep learning models can also deal with various imperfections in the manifolds. 
The work \cite{liu2018direction} evaluates DNN models and is able to predict DoAs using a ULA with an imperfectly known manifold. 
In the work of \cite{papageorgiou2020direction} and  \cite{papageorgiou2020fast}, it is shown that a DNN with a denoising autoencoder performs better than classic subspace methods at low SNR, and can estimate DoA without knowing the exact number of sources
Further, it has been shown that a DNN can deal well with the limited field-of-view problem, outperforming subspace methods at low-angles \cite{xiang2019novel}. These aforementioned techniques however, do not deal well with coherent multipath interference.
However, \cite{wu2019deep} enhances CNN performance with sparse prior information, while \cite{yao2020crnn} shows that a CNN can estimate DoAs of the main source in the presence of coherent interference, although the approach does not estimate the total number of paths or the direction of the multipath.  
Our previous work \cite{yu2020direction} demonstrated that a DNN can estimate the model and even determine if the array is overloaded.

What is needed is a MOE approach that can identify the total number of signals, the number of independent sources, and the multipath associated with those sources.  Thus, in the current work, we propose an efficient DL-based MOE estimation approach by exploiting an advanced CNN architecture known as a Residual CNN or RCNN to precisely classify both the number of signals and sources in the presence of coherent multipath. We further show how it can be used to perform signal association.
We first describe how a lack of understanding of the number signals (including the number of coherent multipath) impacts a  DoA system, and propose a system model for estimating the number of signals, sources and association of signals to sources.  We then define the signal, the coherent interference and the array models. 
Further,  we define three different classification tasks to perform the MOE, and show how the output of these tasks impact the overall system, including temporal smoothing (needed to mitigate the effects of coherent multipath) and signal association. 
Afterwards, we explain the use of the covariance matrix as our input feature, and why the classic methods fail in the presence of coherent multipath. 
Following this, we describe how our proposed RCNN model can deal with this problem. 
In particular we show that in addition to the advanced structure, the use of a weighted loss function which errs on the side of over-estimations improves the overall system performance. 
Finally we provide comprehensive simulation results to verify our methods and compare to existing approaches. 
We provide results demonstrating the computational cost, the impact of field-of-view, and the overloaded array case.

The contributions and results are summarized as follows:
\begin{itemize}
 \item Ours is the first work to apply the RCNN architecture to the problem of multiple label classification, which here includes estimation of three separate labels to determine the number of sources, the number of multipath signals per source and the total number of signals. 
 We show that the RCNN is more powerful than conventional CNN or MLP approaches at extracting key information from the covariance matrix, and even distinguishes detailed information that traditional MOE techniques cannot.

\item Our technique estimates three separate labels to inform additional parameter estimation techniques.  Specifically, we estimate the minimum blocks required for temporal smoothing, the total number of individual paths,  needed in DoA estimation, and the number of uncorrelated sources to facilitate signal association. 
These techniques are fundamental to a DoA-based localization and tracking system and are enabled via our MOE technique.

\item Since, as we later show, overestimation carries a lower performance penalty than underestimation for this problem, we develop a weighted loss function which penalizes underestimation more than overestimation. We show that the overall localization and tracking system is robust to this biased error.

\item We conclude with a comprehensive performance analysis not only of our MOE structure, but of the overall localization and tracking system. 
We investigate several array geometries, independent classification tasks, and multiple deep learning model architectures. 
We provide a discussion of overloaded classification, which occurs when the number of incident signals exceeds the theoretical limit. 
Finally, we provide an examination of the computational cost and compare our approach to the state of the art, including both traditional MOE techniques as well as other machine learning based approaches.  These comparisons show the superiority of the proposed approach.

\end{itemize}

\begin{figure*}[t]
     \begin{center}
     \includegraphics[width=1.00\textwidth]{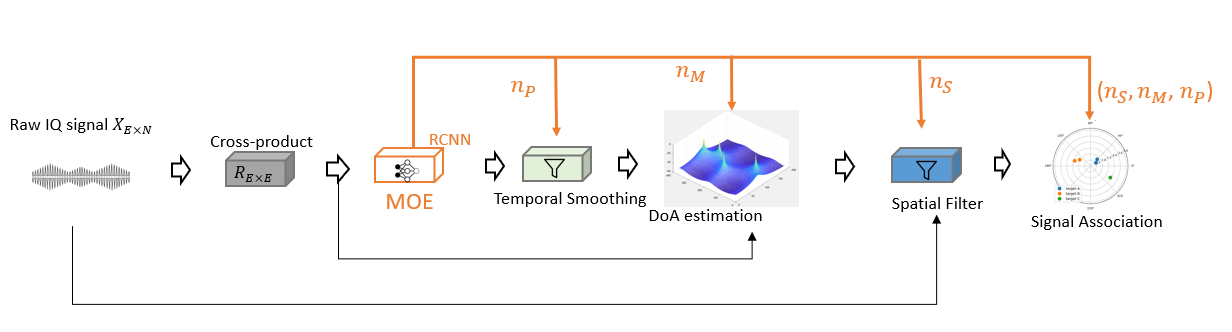}
        \end{center}
    \vspace{-0.2in}
    \caption{Flow diagram  of signal isolation and analysis system: The RCNN provides $n_P$ for temporal smoothing, $n_M$ for MUSIC-based DoA estimation, and $n_{P},n_{M},n_{T}$ for signal association. }
  \label{fig:0_overflow}
      \vspace{-0.15in}
\end{figure*}


\section{Proposed System and Models}

In this section we provide the problem formulation along with the proposed system, the signal and array models, as well as the three classification tasks needed to solve the overall problem.

\subsection{Problem Formation}
We assume the use of an antenna array of $E$ elements that receives $n_M$ total signals coming from $n_S$ sources. 
Note that $n_M \geq n_S$, due to the multipath model we show later. 
We wish to estimate the total number of signals $n_M$, the number of independent sources $n_S$, the DoAs of each signal, and the association of each signal to a source.  
Our overall signal processing flowchart is shown in Fig. \ref{fig:0_overflow}. 
The processing is composed of six blocks:
\begin{enumerate}
    \item \textit{Cross product block}: This pre-processing component converts raw IQ signals into a covariance matrix used by the MOE and DOA-estimation blocks. The covariance matrix $\mathbf{R}_{xx}$ is an estimate of the source covariance and is the feature used by the RCNN.  It is also used by temporal smoothing.
    \item \textit{MOE}: This block estimates the number of signals $n_M$, independent sources $n_S$, and maximum paths per source $n_P$.  The block utilizes an RCNN to accomplish this task as will be described in detail.
    \item \textit{Temporal smoothing}: This block mitigates the impact of coherent multipath to allow DoA estimation.  It uses $n_B$ covariances matrices to estimate a covariance matrix suitable for DoA estimation.  Note that $n_B=n_P+1$ and thus this block relies on the output of the MOE block.  
    \item \textit{DoA estimator}: This block uses a traditional subspace-based method like MUSIC to estimate the DoA of all signals.
    \item \textit{spatial filter}: This block spatially filters the received signal to isolate each received signal and facilitates signal association. 
    \item \textit{signal association}: This block performs signal correlation which identifies the sources associated with each signal.  Again, this relies on the output of the MOE block.
\end{enumerate}

Summarizing the processing, a snapshot of complex port voltages is collected on an electromagnetic array. 
We represent collected data as an $N\times E$ matrix of $N$ complex time samples taken from $E$ ports. 
A cross-product is performed on the data, resulting in an $E\times E$ complex matrix estimate of the covariance of the input. 
The real and imaginary components of this matrix are separated, transforming the dimensionality to an $E\times E\times2$ array for input into the RCNN. 
We feed this image-like object into our proposed RCNN model, and the learning model predicts the multipath state as one of several possible states, shown later in Table \ref{table:MOE}. 
The output of the classifier  includes: 
\begin{enumerate}
    \item An estimate of the maximum number of paths any signal propagates along, $n_P$. This informs the temporal smoothing algorithm. 
    \item An estimate of the total number of signals arriving at the sensor, including multipath copies $n_M$. 
    \item An estimate of the number of \emph{uncorrelated} signals $n_S$. This is equivalent to the total number of sources in the environment. 
\end{enumerate}
Note that the symbols $n_S$, etc. denote the true values. When we intend to discuss the estimate, we write $\hat{n_S}$. Throughout the rest of this work, we assume a MUSIC-like DoA estimator when DoA estimates are required.  After that, we feed the DoAs into a spatial filter to obtain each of the received signals, and then in signal association steps, we color or cluster these angles based on their source.

\subsection{Signal and Array Models}

\begin{figure}[t]
  \centering
  \includegraphics[width=0.50\textwidth]{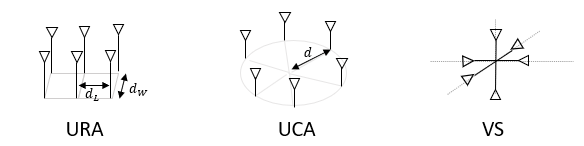}
  \caption{Different antenna arrays considered in this work}
  \label{fig:2_arrays}
\end{figure}

We discuss three types of common arrays as shown in Fig. \ref{fig:2_arrays}: the Uniform Rectangular Array (URA), Uniform Circular Array (UCA), and the Vector Sensor (VS) \cite{haykin1985array}. 
We assume that each array has six elements and can therefore provide DoA estimates for up to 5 sources \cite{nehorai1991vector}.  
Due to the constant number of  elements across arrays, we can describe the received signal for all three array geometries using the same general formula:
\begin{equation}
    \label{eq:sig_model}
    \mathbf{X}(n) = \mathbf{A}(\psi)\mathbf{S}(n) + \mathbf{w}(n)
\end{equation} 
where $\mathbf{X}(n)$ is the $n$th block of samples, i.e., an $N\times E$ matrix of complex signal samples as observed at each element and $\mathbf{w}$ is an $N\times E$ matrix representing additive (temporally and spatially) white Gaussian noise.
As shown in Table. \ref{table:symbol}, $E$ denotes the number of antenna elements 
and $N$ is the number of samples captured during the collection interval (i.e., the block size).
The matrix $\mathbf{A}$ is of size $E \times n_M$ and is comprised of the array steering vectors for each received signal, i.e.,  $\mathbf{A} = [\mathbf{a}(\psi_1),\mathbf{a}(\psi_2),...\mathbf{a}(\psi_{n_M})]$ and the vector $\psi_k$ is a vector containing the parameters of the steering vectors, i.e.,  $\psi_k = (\theta_k, \phi_k)^T$ where  $\theta_k$ is the azimuth angle and $\phi_k$ is the elevation angle.  Each type of array has a different expression of the steering vector  $\mathbf{a}(\psi_k)$. 
For a URA, the steering vector is

\begin{equation}
\mathbf{a}_{URA}(\psi_k) = [a_{1,1}(\psi_k),a_{1,2}(\psi_k),\dots ,a_{L,W}(\psi_k)], 
\end{equation}
where
\begin{equation}
a_{i,j}(\psi_k)  = e^{2\pi j \sin \theta_k (\cos\phi_k \frac{i d_W}{\lambda} + \sin \phi_k  \frac{j d_L}{\lambda}) }    
\end{equation}
The variables $L$, $d_L$, $W$,  $d_W$ represent the horizontal antenna number, horizontal antenna spacing, vertical antenna number and  vertical antenna spacing respectively. 

The UCA steering vector is expressed as Eq. (\ref{eq:UCA_steer}).  
\begin{align}
    \label{eq:UCA_steer}
    \mathbf{a}_{UCA}(\psi_k) = 
        & [e^{2 \pi j (d/\lambda) sin(\theta)cos(\phi  ) },\dots \\
        & e^{2 \pi j (d/\lambda) sin(\theta)cos(\phi -  2\pi i/E) },\dots \\
        & e^{2 \pi j (d/\lambda) sin(\theta)cos(\phi - 2\pi (E-1)/E)} ]  
\end{align} 
Finally, the vector sensor steering vector is shown in Eq. (\ref{eq:VS_steer}). 
\begin{equation}
\label{eq:VS_steer}
    \mathbf{a}_{VS}(\psi_k)=
        \begin{bmatrix}
          \cos \theta_k \cos \phi_k  & -\sin\phi_k \\
          \cos \theta_k \sin \phi_k  & \cos\phi_k \\
          - \sin \theta_k  & 0\\
          - \sin \theta_k  & -\cos \theta_k \cos \phi_k\\
          \cos \phi_k  & -\cos \theta_k \sin \phi_k\\
          0 & \sin \theta_k
    \end{bmatrix}
    \begin{bmatrix}
          \sin \gamma_k e^{j\eta_k} \\
          \cos \gamma_k
    \end{bmatrix} 
\end{equation}
Note the polarization angles $\gamma_k$ and $\eta_k$ are outside of our interest here and are set as fixed values. Several vector sensors, each composed of six elements, can also be arranged  to detect more than six sources, but is outside of our focus in this work.
For all arrays, the domain of the elevation and azimuth are $\theta_k \in [0,\pi)$ and $\phi_k \in [0,2\pi)$ respectively. 
When we describe a general model without conditioning on the received array we write $\mathbf{a}(\psi_k)$.

\begin{table}
\caption{Notation  }
\label{table:symbol}
\begin{center}
\begin{tabular}{c|c}
\hline
 Symbols & Meaning\\
\hline
$n_S$ & Number of independent sources\\
\hline
$n_M$ & Total number of received paths\\
\hline
$n_P$  & Maximum number of paths for any source\\
\hline
$n_{SMP}$ & Scalar index mapped to a unique $(n_S,n_M,n_P)$ \\
\hline
$E$ & Number of antenna elements \\
\hline
$N$ & Number of signal samples per snapshot\\
\hline
$n_B$ & Number of observation blocks \\
\hline
$t_{coh}$ & Coherence time \\
\hline
$\mathbf{R}$ & The temporally smoothed covariance matrix estimate\\
\hline
$\mathbf{R}_b$ & The covariance matrix estimate for a single received signal block\\
\hline
$\mathbf{a}$ & steering matrix \\
\hline
$h$ & channel matrix \\
\hline
$\theta$ & azimuth angle, range in  $  (0, 2\pi)$ \\
\hline
$\phi$ & elevation angle, range in  $  (0, \pi)$ \\
\hline
\end{tabular}
\end{center}
\end{table}

Multipath propagation is the phenomenon where a signal transmitted from an emitter travels along several paths of different lengths before being received from potentially different directions and at potentially different times. 
Depending on the time and angle difference between multipath arrivals, multipath can result in  spatial correlation and/or temporal fading in the received signal, since the multipath signal will arrive with multiple different phases or angles. 

This model assumes that the channel exhibits Rayleigh fading in time and that all spatially distinct paths fade independently. 
For an individual spatial path (i.e., collection of signals coming from the same direction), the fading at two time instants can be regarded as uncorrelated if the time interval between these two instants is larger than the coherence time $t_{coh}=c /\left(v f_{c}\right)$, where $c$, $v$, and $f_{c}$ are the speed of light, the source's relative velocity and the signal carrier frequency, respectively. 
For a discrete electromagnetic source $s$, with $s\leq n_S$, we say that the signal is received on $n_p^{(s)}$ paths. 
A critical value for removing the effects of multipath is the maximum number of paths any single source propagates along. 
We can find this value as Eq. (\ref{eq:max_paths}). 
\begin{equation}
\label{eq:max_paths}
    n_P = \max n_p^{(s)}
\end{equation}

We will describe a \emph{block-fading channel} here. 
The block fading channel model is defined with fading which varies between blocks of received signal samples. 

Importantly, the block-fading channel assumes that fading is constant throughout the duration of each block. 

Denote the signal received from the $s^{th}$ source as $X^{(s)}$. 
Then, following the block fading channel, represent the $b^{th}\leq B$ \emph{block} of $X^{(s)}$ as $X_b^{(s)}$. 
Further, following the signal model in Eq. (\ref{eq:sig_model}) and incorporating the multiple paths propagated by source $s$, we can write
\begin{equation}
\label{eq:single-signal-multipath-sig-model}
    X_b^{(s)}(n) = \sum_{p=1}^{n_p^{(s)}} h_{s,p} \mathbf{a}(\psi_{s,p}) \mathbf{S}_{s}(n+\delta_{s,p})
\end{equation}
Note several things: 
\begin{itemize}
    \item $\mathbf{h}_{s,p}$ represents a \emph{scalar} complex phase shift. 
    \item $\delta_{s,p}$ denotes the path delay relative to line-of-sight experienced by the $p^{th}$ copy of the $s^{th}$ source. 
    \item The time index $n\leq N_B$ denotes the sample length of each block, and $N_B$ is constant from block to block. 
\end{itemize}

Now, with an understanding of the effects of multipath on the single-signal model, we can proceed to describe the multi-signal, multipath case Eq. (\ref{eq:one_block_sig}). 
\begin{equation}
\label{eq:one_block_sig}
    X_b = \sum_{s=1}^{n_S} X_b^{(s)} + \mathbf{w}
\end{equation}
Note that we remove the dependence on $n$. 
The complex additive white Gaussian noise $\mathbf{w}\sim\mathcal{N}(0,\sigma_w^2)$ matrix is assumed to be the appropriate size. 
It is assumed that the transmitting sources are continually transmitting, and only $B$ blocks are collected by the receiving system. 
Finally, it is up to the receiver to determine how many blocks must be received to facilitate decoherence via temporal smoothing. 
Before this can be determined, however, the system must first perform MOE.

\begin{table*}[t]
\centering
\caption{Label Notation Used for MOE classifier} 
\begin{tabular}{c  c  c  c c }
\hline
 5-category $n_M$ & 9-category & 18-category $n_{SMP}$  & Notation & $ (n_S, n_M, n_P) $ \\
\hline
$1$ &   $1$ & $1$ & $(A)$  & (1,1,1) \\
\hline
$2$ &   $2$ & $2$ & $(A,A)$ & (1,2,2)  \\
$2$ &   $3$ & $3$ & $(A,B)$ & (2,2,1)   \\
\hline
$3$ &   $4$ & $4$ & $(A,A,A)$  & (1,3,3)  \\
$3$ &   $5$ & $5$ & $(A,A,B)$  & (2,3,2)   \\
$3$ &   $5$ & $6$ & $(A,B,C)$  & (3,3,1)  \\
\hline
$4$ &   $6$ & $7$ & $(A,A,A,A)$  & (1,4,4)  \\
$4$ &   $7$ & $8$ & $(A,A,A,B)$   & (2,4,3) \\
$4$ &   $7$ & $9$ & $(A,A,B,B)$   & (2,4,2) \\
$4$ &  $7$ & $10$ & $(A,A,B,C)$   & (3,4,2)  \\
$4$ &  $7$ & $11$ & $(A,B,C,D)$   & (4,4,1)   \\
\hline
$5$ &  $8$ & $12$ & $(A,A,A,A,A)$  & (1,5,5)  \\
$5$ &  $9$ & $13$ & $(A,A,A,A,B)$ & (2,5,4)  \\
$5$ &  $9$ & $14$ & $(A,A,A,B,B)$ & (2,5,3)  \\
$5$ &  $9$ & $15$ & $(A,A,A,B,C)$  & (3,5,3) \\
$5$ &  $9$ & $16$ & $(A,A,B,B,C)$ & (3,5,2)  \\
$5$ &  $9$ & $17$ & $(A,A,B,C,D)$ & (4,5,2)  \\
$5$ &  $9$ & $18$ & $(A,B,C,D,E)$  & (5,5,1)  \\
\hline
\end{tabular}
\label{table:MOE}
\end{table*}

\subsection{Classification Tasks}

Our classifier model performs three separate classification tasks: 
\begin{enumerate}
    \item Five-category classification, for determining the quantity of total \emph{paths} incident on the sensor. 
    \item Nine-category classification, which determines not only the number of total paths, but also  whether or not multipath is from a single source or multiple sources. For example, as shown in Table \ref{table:MOE} class 4 indicates that there are three signals present but only one source, whereas class 5 indicates that there are three signals present and multiple sources.
    \item 18-category classification, which estimates the exact multipath state (i.e., the number of sources and the number of multipath for each source). 
\end{enumerate}
When we refer to multipath state, we mean the tuple $(n_S, n_M, n_P) \in [1,5]^3$. 
Each possible signal model can be represented as a unique element in this space. 
Our classifier estimates a unique scalar label $n_{SMP}$  which is mapped to the tuple label space according to the function $\mathbb{G}$, Eq. (\ref{eq:mapping_fn}). 
\begin{equation}
    \label{eq:mapping_fn}
    n_{SMP} = \mathbb{G}(n_S, n_M, n_P)
\end{equation}
Put another way, once $n_{SMP}$ is determined, we can determine $n_S = \mathbb{H}^S(n_{SMP})$,  $n_M = \mathbb{H}^M(n_{SMP})$ and $n_P = \mathbb{H}^P(n_{SMP})$ using the one-to-one mapping from the Table \ref{table:MOE}.
Note that there are some elements $(n_S, n_M, n_P)$ which are not represented in the possible signal model space. 
Therefore, call the space of valid tuples $\mathbb{L}$. Note that if we remove any label from $(n_S,n_M,n_P)\in\mathbb{L}$, we lose the bijection in $\mathbb{G}$. 
In other words, we require \emph{all three} labels in the tuple to have a one-to-one mapping between $(n_S,n_M,n_P)$ and $n_{SMP}$. 
Since the space is reasonably small, we provide a complete list in Table \ref{table:MOE}. 

Rather than use the entire received signal as our feature set for estimation, we will reduce it to the covariance matrix calculated as $\mathbf{R}_b = \frac{1}{N_B}X_b X_b^H$, where $(\cdot)^H$ is the complex conjugate transpose. 
Therefore the MOE function works on $\mathbf{R}_b$, conditioned on a specific array geometry $\mathbf{a}$. 
\begin{equation}
    n_{SMP} = f_{MOE|\mathbf{a}}(\mathbf{R}_b) 
\end{equation}

Finally, when we discuss an \emph{estimate} of a label $n_X$, we will denote $\hat{n}_X$. For example, $\mathbb{G}(\hat{n}_{SMP}) = (\hat{n}_S, \hat{n}_M, \hat{n}_P)$.

\subsection{Temporal Smoothing}
Temporal Smoothing is a technique used to suppress coherent signals. 
It is shown in \cite{gu2003performance} that by averaging over covariance matrices, the effects of Rayleigh fading can be mitigated. 
Temporal smoothing requires collection over several independent sensing periods, called \emph{blocks}, separated by a time greater than $t_{coh}$. 
Therefore, we choose a particular subset of blocks with \emph{odd} indices, to ensure separation of greater than $t_{coh}$. 
Also note that in order to this technique requires the use of $B\geq n_P+1$ blocks. 
In practice, the system selects $B = \hat{n_P}+1$. 
Since the transmitting sources are assumed to be constantly emitting, the system is able to select $B$ without constraints, thus it is not an estimate. 
We can write the final temporally smoothed covariance matrix $\mathbf{R}$ as Eq. (\ref{eq:TS_cov}). 
\begin{equation}
\label{eq:TS_cov}
    \mathbf{R} = \frac{1}{B} \sum_{b=0}^{B-1} \mathbf{R}_{2b+1}
\end{equation}

In practice, $\hat{n}_P$ is formed using $R_1$, which informs $B$. 
In the event that a multipath channel is present and the model order is underestimated resulting in fewer blocks than required, DoA estimators tend to fail due to the remaining correlation. 
However, \emph{overestimation} of $n_P$ and using a greater $B$ than necessary results in a greater SNR due to the increased collection time. 
This of course comes at the cost of greater computation and time per DoA estimate, so it is not generally desirable to overestimate $n_P$. 

\begin{figure}[t]
  \centering
  \includegraphics[width=0.40\textwidth]{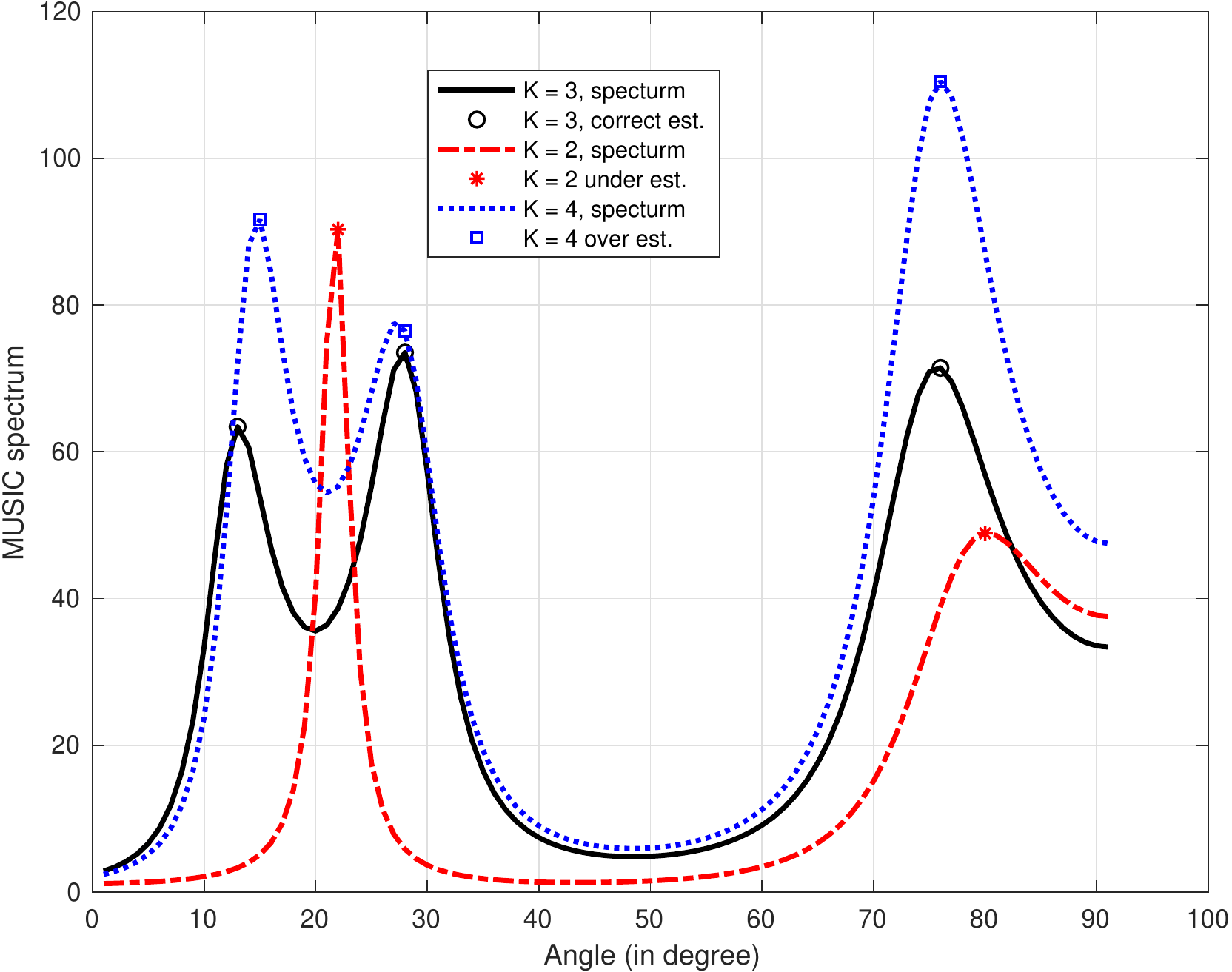}
  \caption{Illustration of underestimation and overestimation on MUSIC Spectrum}
  \label{fig:11_underEst}
\end{figure}

These effects are demonstrated in Fig. \ref{fig:11_underEst}. 
Three curves are shown: for underestimated model order of 2, correctly estimated model order of 3, and for overestimated model order of 4. 
There are three paths present. 
When the model order is underestimated, the DoA estimator tends to average out the spatial diversities. 
The correct model order results in correct estimation. 
Then, when the model order is overestimated, the peaks in the spatial spectrum become more pronounced.

\begin{figure}[t]
  \centering
  \includegraphics[width=0.40\textwidth]{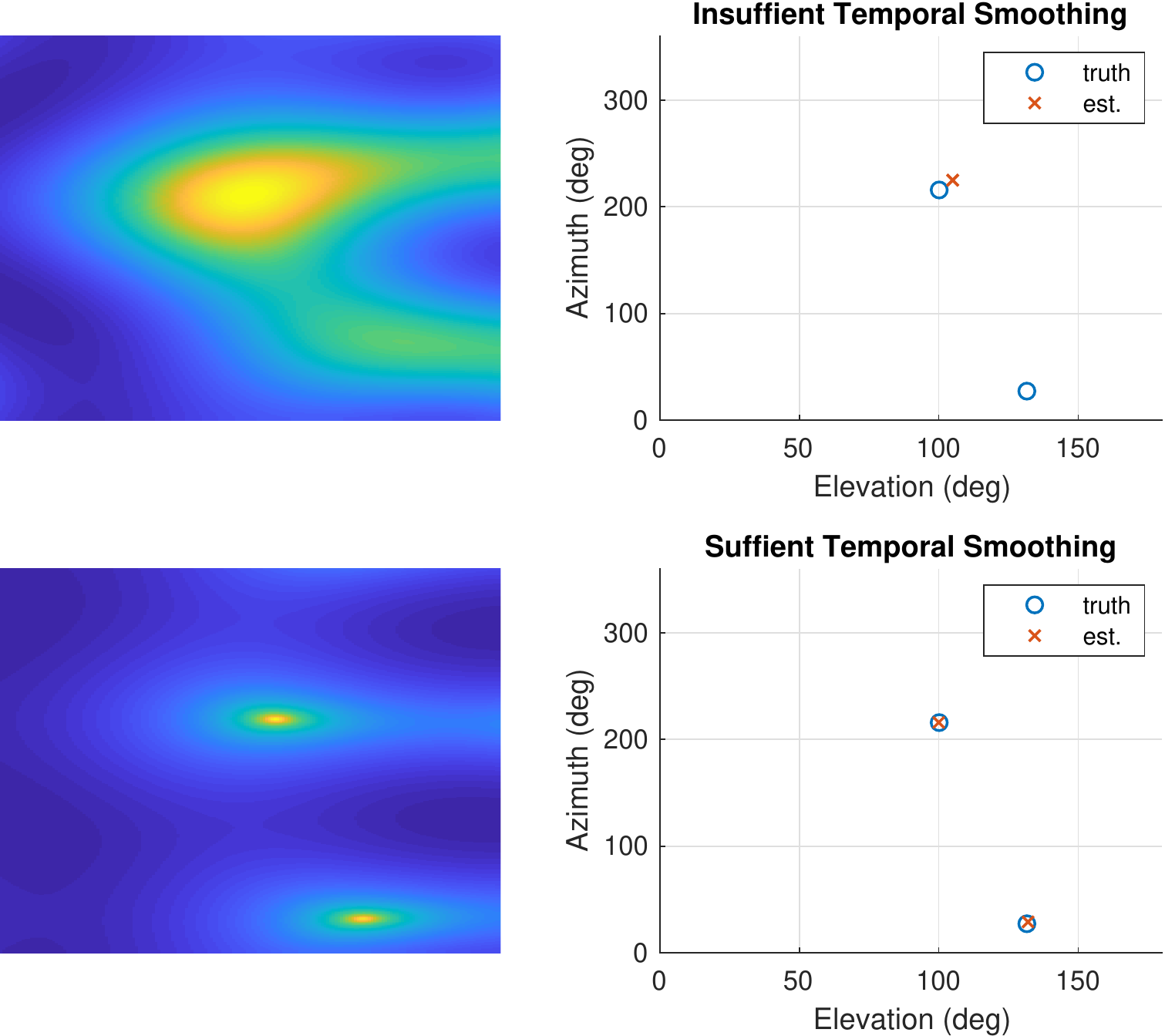}
  \caption{Illustration of 2D MUSIC spectrum after temporal smoothing with underestimated (top) or overestimated (bottom) number of total paths}
  \label{fig:12_ts}
\end{figure}

An additional effect occurs in the actual DoA estimator. 
A typical input to MUSIC-type estimators is the model order. 
When this value is underestimated, these estimators tend to fail. 
However, when this value is instead overestimated, the impact is less severe. Fig. \ref{fig:12_ts} illustrates this. 
The top plot (blue) represents the MUSIC spectrum when the model order is underestimated. As discussed above, overestimation tends to average out the spatial diversity between the paths. 
The bottom plot (black) shows the MUSIC spectrum using the correct MOE, while the middle plot (red) shows the impact of MOE underestimation on the MUSIC spectrum. In conclusion, since estimation error cannot be fully eliminated, we bias towards overestimation, since temporal smoothing will still be beneficial. Moreover, overestimation allows DoA estimation to be reasonably good.


\subsection{Signal Association}

\begin{figure}[t]
  \centering
  \includegraphics[width=0.25\textwidth]{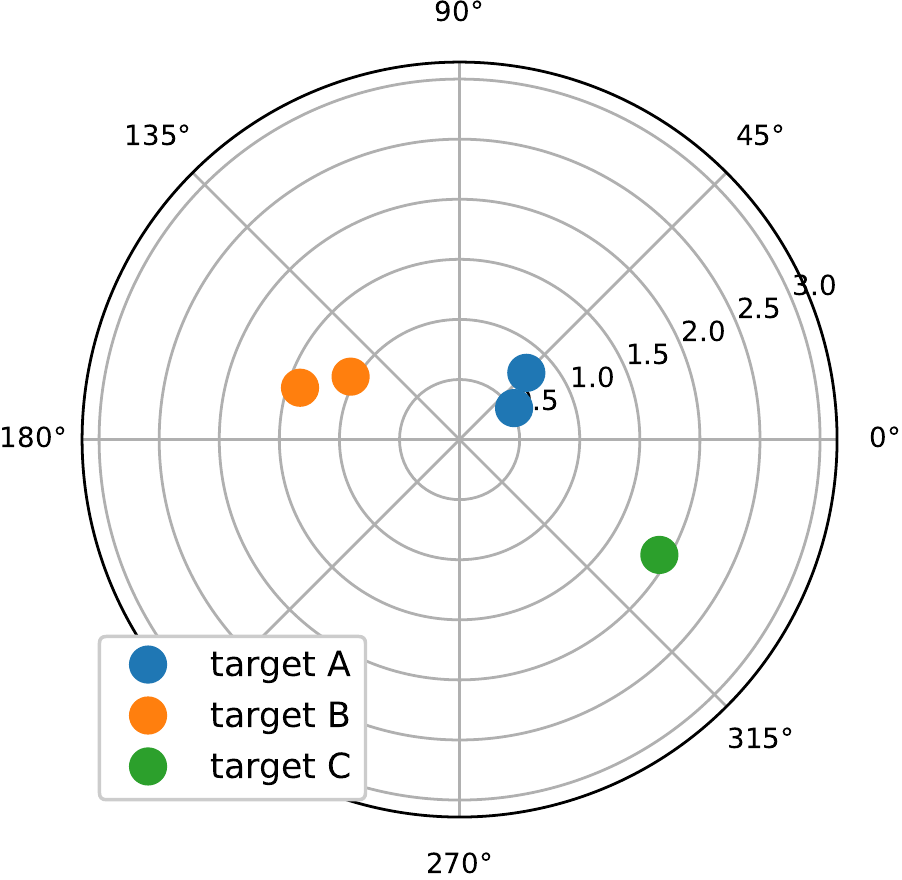}
  \caption{Illustration of signal association, angles with same color mean their signal is correlated, and this  is notated as 16th case (A,A,B,B,C) }
  \label{fig:sa}
\end{figure}

\begin{figure*}
     \begin{center}
     \includegraphics[width=0.80\textwidth]{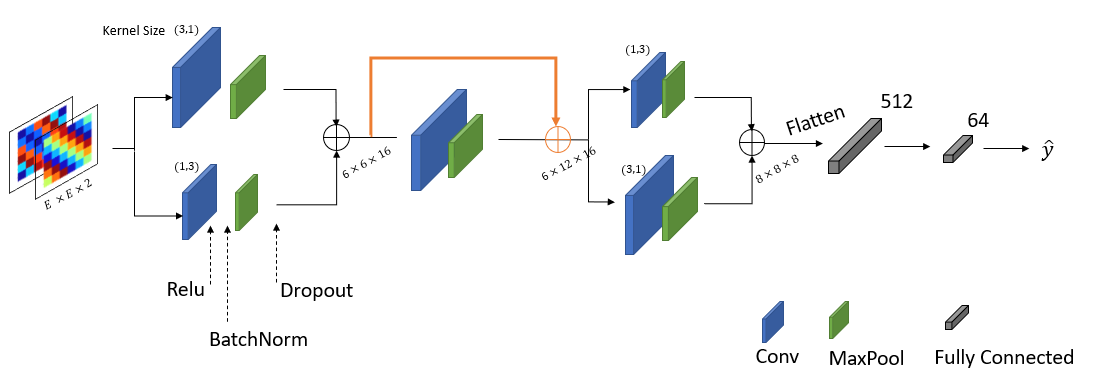}
        \end{center}
    \vspace{-0.2in}
    \caption{Residual Convolution Neural Network structure, with grouped asymmetric convolution kernels and residual links}
  \label{fig:2_RCNN}
      \vspace{-0.15in}
\end{figure*}

So far, we have described the classification model used to estimate the useful MOE parameters such as $n_P$, which informs temporal smoothing and thus impacts DoA estimation. 
Once the signal directions have been estimated, we need some technique to assign each path to a  source. 
Recall that when multipath is present, more than one [path/]direction may need to be associated to a single source. 
This is accomplished via signal association; by analyzing the information coming from each direction, clusters can be formed.

We provide an example in Fig. \ref{fig:sa} to illustrate the usefulness of signal association. 
In this example, the true label is $(n_S,n_M,n_P) = [3,5,2]$. 
So, we should expect to see three different uncorrelated sources, five different directions, and a maximum of two different paths for any single signal. 
In this figure, the radial axis denotes elevation, and the angular axis denotes azimuth. 
After signal association, we color each detection according to its likely source. 

 In order to determine which signals are associated with which source, we first need to extract the signals from each direction.  This is done via spatial filtering using the following steps: 
\begin{enumerate}
    \item Estimate the steering vector $\hat{a}_i$ for the  $i$th direction as $\hat{a}_i = a(\hat{\theta_i})$, where $\hat{\theta}_i$ is the DoA estimate for the $i$th direction. 
    \item Estimate $\hat{\mathbf{W}}_i=\mathbf{R}^{-1}\hat{a}_i$, where $\mathbf{R}^{-1}$ represents the pseudo-inverse of the received signal covariance matrix $\mathbf{R}$. 
    As compared to the standard inverse, the pseudo-inverse is guaranteed to exist. 
    \item Reconstruct the signal from the $i$th direction as $\hat{s}_i(t) = \hat{\mathbf{W}}_i x(t)$. 
\end{enumerate}

Once spatial filtering is complete, we perform signal association by  correlating each pair of source signals $\hat{s}_i, \hat{s}_j$ as $c_{i,j}=\hat{s}_i(t)\otimes\hat{s}_j(t)$. Pairs that have high correlation are likely to have originated from the same source.

The computational complexity of this signal association scales as $O({n^2_M})$, when each signal is correlated against every other signal. Note that $O$ is the Bachmann–Landau big-O notation. 
This algorithm is shown fully in Algorithm \ref{alg:sa_greedy}. 
In short, this greedy algorithm takes $n_M$ as an input, and partitions the detected signals into sets of likely correlated signals. 
The number of these sets is dictated by $n_S$, and the largest set will contain $n_P$ elements.

However, we can do better than the above ``greedy'' in terms of complexity. 
Instead of brute-force correlating every pair of signals, we can narrow down the search range as soon as sets reach $n_P$ elements. 
This lower complexity algorithm is shown in Algorithm \ref{alg:sa}.

\begin{algorithm}
\caption{Greedy signal association algorithm}\label{alg:sa_greedy}
\begin{algorithmic}
\Require $n_M$
\State init null sets $C_1,\dots, C_{n_M}$
\State init $S \gets \{ s_1,\dots, s_{n_M}\} $
\For{$C_i \in \{ C_1,\dots, C_{n_M} \}$} 
\For{$s_j \in S$} 
\If{ $cor(s_i, C_i(1) )$ is True or $C_i$ is null }
    \State $C_i \gets C_i \cup s_j $ \Comment{append $s_j$ to set $C_i$}
    \State $S \gets S ~\backslash~ s_j$  \Comment{remove $s_j$ out of set $S$}
\EndIf
\EndFor
\EndFor
\end{algorithmic}
\end{algorithm}

\begin{algorithm}
\caption{Enhanced signal association algorithm}\label{alg:sa}
\begin{algorithmic}
\Require $ n_S, n_M, n_P $
\State init null sets $C_1,\dots, C_{n_S}$
\State init $S \gets \{ s_1,\dots, s_{n_M}\} $
\For{$C_i \in \{ C_1,\dots, C_{n_S}\}$}
\For{$s_j \in S$} 
\If{ $cor(s_i, C_i(1) )$ is True or $C_i$ is null }
    \State $C_i \gets C_i \cup s_j $ \Comment{append $s_j$ to set $C_i$}
    \State $S \gets S ~\backslash~ s_j$  \Comment{remove $s_j$ out of set $S$}
\EndIf
\If{ len($C_i$) equal $n_P$ }
\State \textbf{break} \Comment{set $C_i$ is full}
\EndIf
\EndFor
\EndFor
\end{algorithmic}
\end{algorithm}

\subsection{MDL and AIC MOE Classifiers}

MDL and AIC \cite{wax1985detection} are among the most widely used methods to estimate the number of signals in a received set of array samples.  
The approaches are based on the fact that the maximum number of signals $n_M$ is equal to the number of elements $E$ minus the number of noise eigenvalues in the covariance matrix. 
Due to the fact that we must use a finite number of samples to estimate the covariance matrix, the noise eigenvalues will not be not equal.  
However, they will be close to each other. 
Such closeness of the eigenvalues can help the estimation. 
Assuming the covariance matrix comes from $N$ snapshots and $d$ signals, then the ratio of their \textit{geometric mean} to their \textit{arithmetic mean} is a measure of closeness of the noise eigenvalues. 
\begin{equation}
    L(d) = -E(N-1)log{\frac{(\prod_{n=d+1}^N \lambda_n)^{1/(N-d)}}{\frac{1}{N-d}\sum_{n=d+1}^{n}\lambda_n}}.
\end{equation}
Based on this measure of closeness, there are two information theoretic criteria,
\begin{equation}
    AIC(d) = L(d) + d(2N-1),
\end{equation}
\begin{equation}
    MDL(d) = L(d) + \frac{1}{2}d(2N-1)logK,
\end{equation}

The index with their minimum value is the estimated number of signals, $n_M = \arg\min_{d} AIC(d)$ or  $n_M = \arg\min_{d} MDL(d)$. Reference \cite{wax1985detection} claims that the MDL approach results in unbiased estimates, while the AIC approach yields biased estimates. Nevertheless, neither of the approaches is useful in the presence of coherent multipath. 
Specifically, we will show later that the approaches fail to estimate the number of signals. To this end, a supervised learning method is proposed to estimate models even with coherent interference.

\section{Residual Neural Network Classifier}

Up to this point, every operation in our signal processing chain requires the received signal covariance matrix $\mathbf{R}$. 
Thus, we choose to continue using $\mathbf{R}$ as the input feature to our machine learning model. 
This is a useful feature to use, since it is of constant dimension for a given array, while the received sample length may vary. 
In addition, it was shown in \cite{yu2020direction} that other feature extraction methods like PCA (Principle Component Analysis) cannot guarantee the retention of the information needed for MOE, while the covariance matrix can.   
In other words, the covariance matrix was shown to be a more efficient data compression tool. 
Further, the covariance matrix suppresses zero-mean spatially white noise sources such as Additive White Gaussian Noise (AWGN). 
As the snapshot length increases, this means that the SNR increases as well. 
A general relationship between these two is that the SNR increases by 3dB for every doubling in snapshot size.

Once $\mathbf{R}$ is calculated, we separate it into real and imaginary parts to form an $E\times E\times 2$ image-like array. 
We choose the Cartesian representation of complex numbers instead of the polar form, due to greater compatibility with loss functions as shown in \cite{yu2020direction}.

In addition to employing conventional MLP and CNN structures, we adapt two other techniques in building the neural network: 
\begin{enumerate}
    \item The grouped asymmetric kernel \cite{ding2019acnet} is used in the CNN layer (known as  ACNet), which strengthens the kernel selection function and is robust to layer parameters. It has been shown in many previous works \cite{tian2021asymmetric}\cite{wang2019adaptively} that this performs better than a single wide CNN layer. 
    \item The residual link \cite{he2016deep} is used to provide high resolution learning with deeper stacked layers. Since the CNN suffers from the vanishing gradient problem and degrades performance with a higher number of  stacked layers, use of the  residual link allows a layer to directly learn the residual of the previous layer's output, which enhances the minor profile.  
\end{enumerate}
  Therefore we format our RCNN structure as shown in Fig. \ref{fig:2_RCNN}. 
The $E\times E\times 2$ input goes through two parallel CNN layers, followed by a ReLu activation layer used for helps to prevent the exponential growth in the computation, a batch normalization that makes the optimization landscape significantly smoother by using normalizing smaller mini batch, a maxpool layer to progressively reduce the spatial size of the latent space, and dropout layers which prevent a model from overfitting and robust to outliers. 
Since the use of these latter three are very common \cite{ioannou2017deep}, we skip their visualization. 
Note that the symmetric kernel on each branch is used to aggregate the adjacent samples along both horizontal and vertical axes to provide similar learning efficiencies \cite{iandola2016squeezenet}. 
These two branches are later merged into a $6\times 6\times 16$ space, which is fed directly into another CNN layer. 
In particular, \emph{residual connectivity} is implemented via a direct path parallel to this CNN layer, summed to the CNN output via element-wise addition. 
This regulates the learning as shown in \cite{he2016deep}. 
This technique helps avoid back-propagation gradient degradation and performance saturation in sequentially stacked layers. 
The next layer consists of two grouped \emph{conv} layers with asymmetric kernel sizes. 
The final processing consists of flattening into one dimension. 
We use a version of cross entropy weighted for classification problems, and the Adam optimizer.

Typically, DNNs use cross-entropy loss when used for supervised learning. 
In its typical form, Eq. \ref{eq:cross_entropy},the cross-entropy loss is symmetric - the same loss is observed for picking category A when B is true and for picking category B when A is true.  
In this form, $p_i \in (0,1)$ represents the ground truth label probability, and $\hat{p}_i \in (0,1)$ represents the estimated (softmax) probability. 
\begin{equation}
\label{eq:cross_entropy}
    L_{loss}=-\sum_{i=1}^{n_{SMP}} p_i \log(\hat{p}_i)
\end{equation}
However, in our MOE application, we have found that estimators biased towards overestimation will outperform those which are unbiased. 
This holds true for both $n_P$ and $n_M$. 
In order to ensure that our estimator becomes biased in this fashion, we propose the weighted cross-entropy loss function shown in Eq. \ref{eq:custom_cross_entropy}.
\begin{align}
\label{eq:custom_cross_entropy}
\begin{split}
    L_{loss} = 
        & - \frac{1}{2} \sum_{i=1}^{n_{SMP}}  {w^i_1} p_i \log(\hat{p_i})   - \frac{1}{2}\sum_{i=1}^{n_{SMP}}   {w^i_2} p_i \log(\hat{p_i}) \\
        = & - \frac{1}{2} \sum_{i=1}^{n_{SMP}}  { e^{K_1(n^i_M - \hat{n}^i_M)} } p_i \log(\hat{p_i})  \\ 
        & - \frac{1}{2}\sum_{i=1}^{n_{SMP}}   {e^{K_2(n^i_P   -  \hat{n}^i_P)} } p_i \log(\hat{p_i}) 
\end{split}
\end{align}
In this weighted form, 
\begin{align}
    \hat{n}_{TMP}^i &= \mathbb{G}(\hat{n}_T^i, \hat{n}_M^i, \hat{n}_P^i)\\
    w^i_1         &= e^{K_1(n^i_M - \hat{n}^i_M)}\\
    w^i_2         &= e^{K_2(n^i_M - \hat{n}^i_M)}
\end{align}
and we select $K_2 >> K_1 > 1$. 
Note also that when $\hat{n}_M^i < n_M^i$, we have $e^{K_1(n_M^i - \hat{n}_M^i)}<1$ and therefore larger penalties for underestimation. 
Further, when $\hat{n}_M^i > n_M^i$, we have $e^{K_1(n_M^i - \hat{n}_M^i)}>1$ and thus we assess smaller penalties for overestimation. While the model is biased, it will still prefer correct classification.  
This is because  the overall loss for \emph{correct} classification is much less than that for overestimation since when  $\hat{n}_M^i = n_M^i$, then $ \lim_{p_i  \to 1}p_i \log(\hat{p_i}) = 0$.

\section{Performance Evaluation}

\begin{figure*}
 \centering
 \subfigure{
 \includegraphics[width=0.22\linewidth]{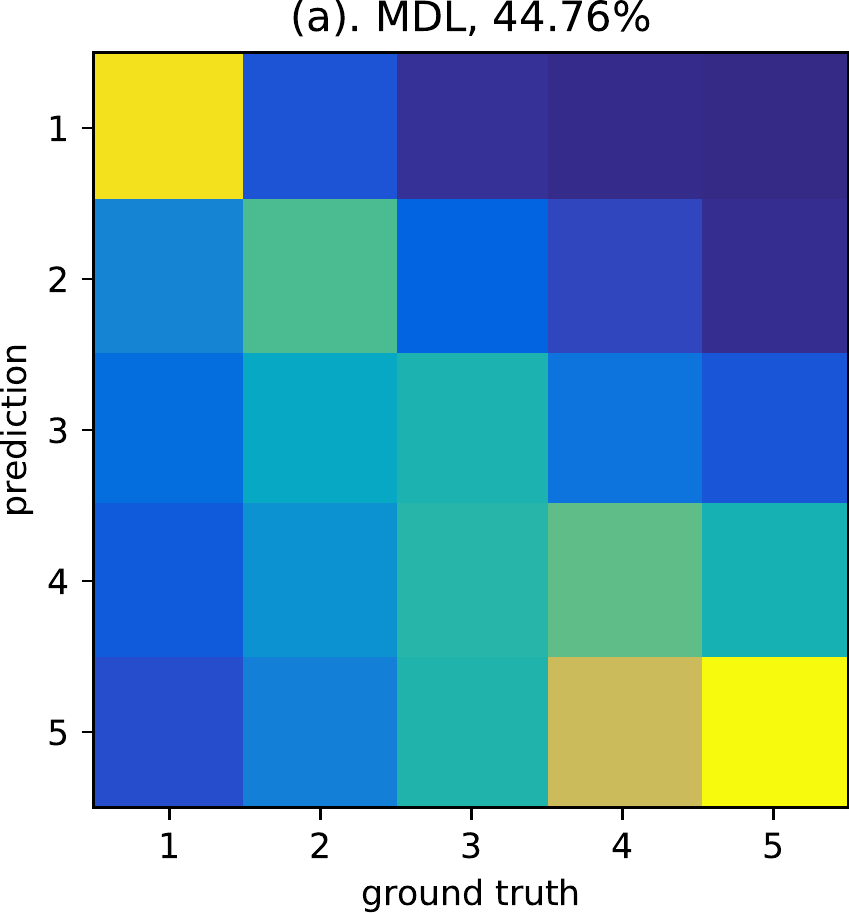}
 }
 \subfigure{
 \includegraphics[width=0.22\linewidth]{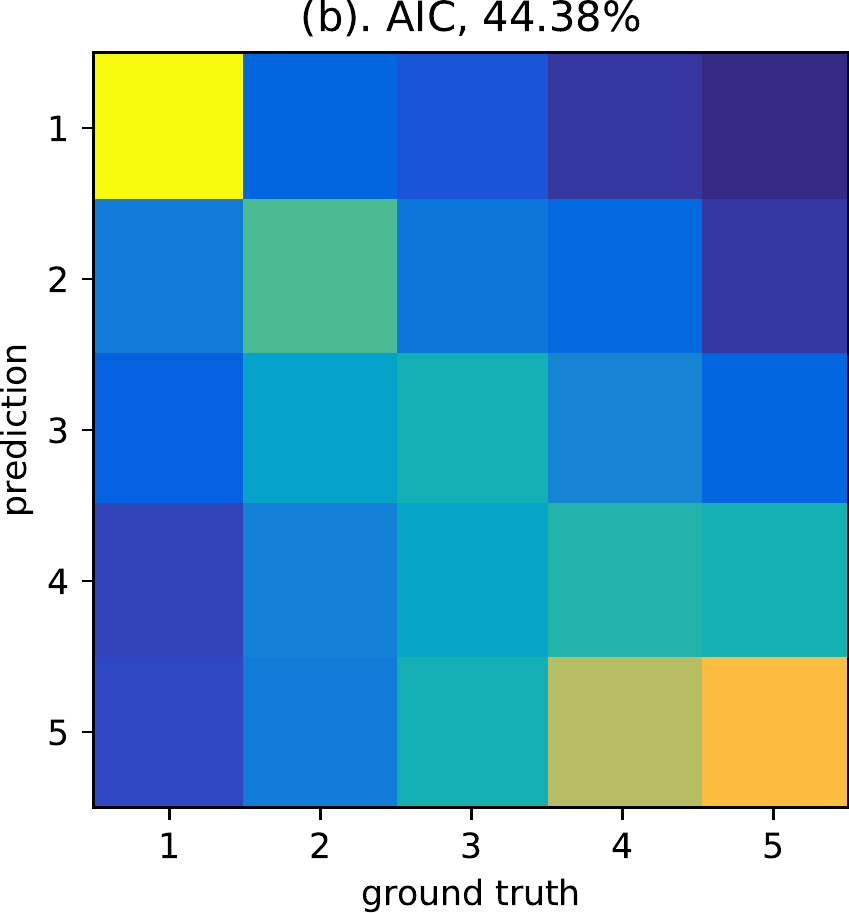}
 } 
 \subfigure{
 \includegraphics[width=0.22\linewidth]{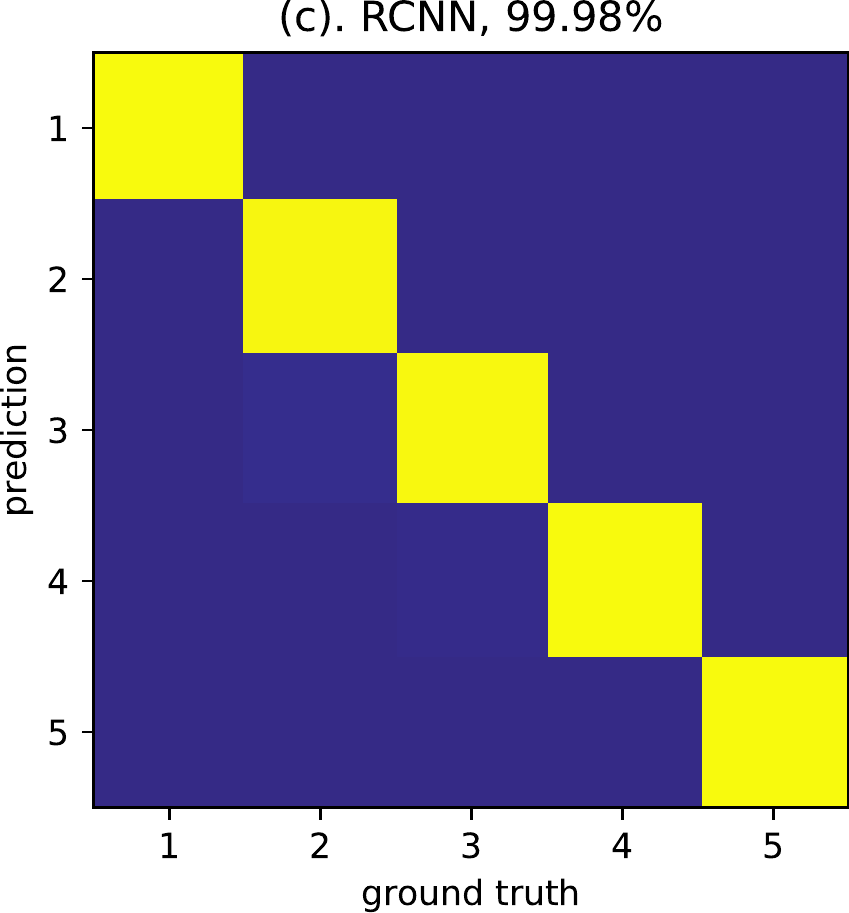}
 } 
  \subfigure{
 \includegraphics[width=0.22\linewidth]{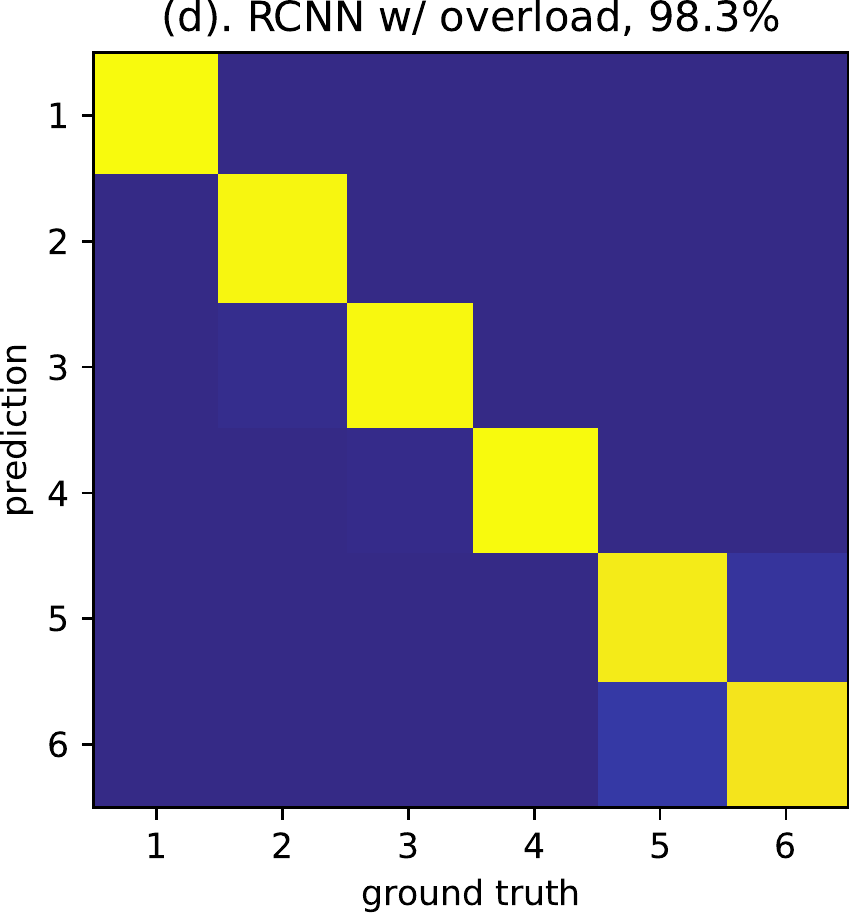}
 } 
 \\
 \subfigure{
 \includegraphics[width=0.22\linewidth]{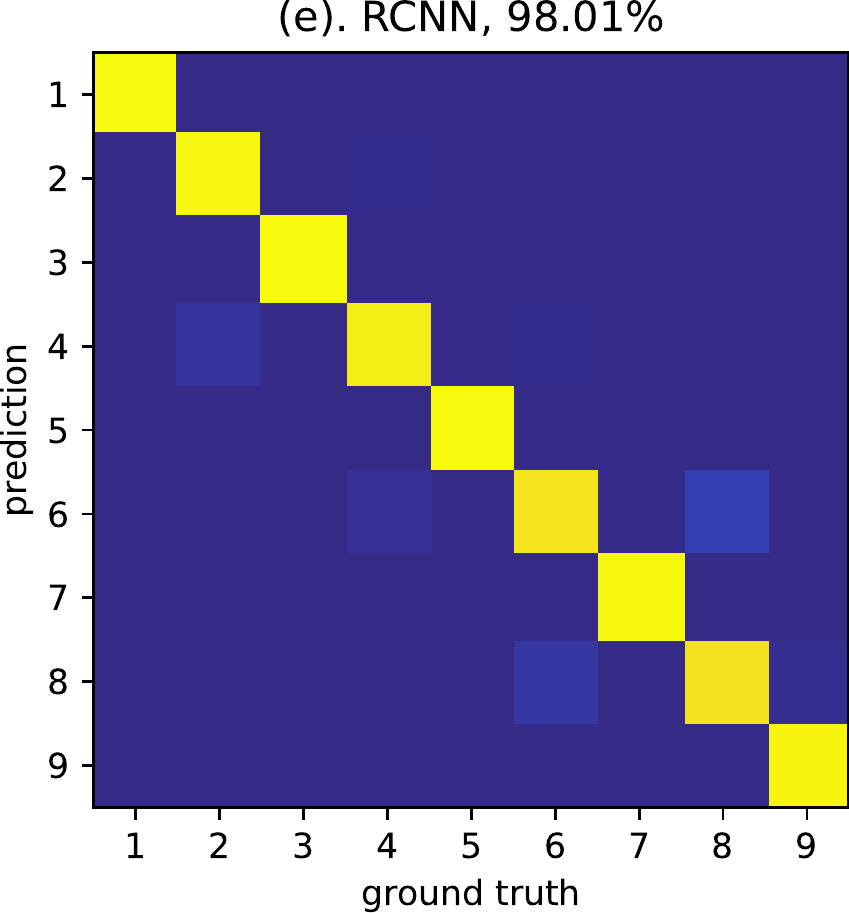}
 }
 \subfigure{
 \includegraphics[width=0.22\linewidth]{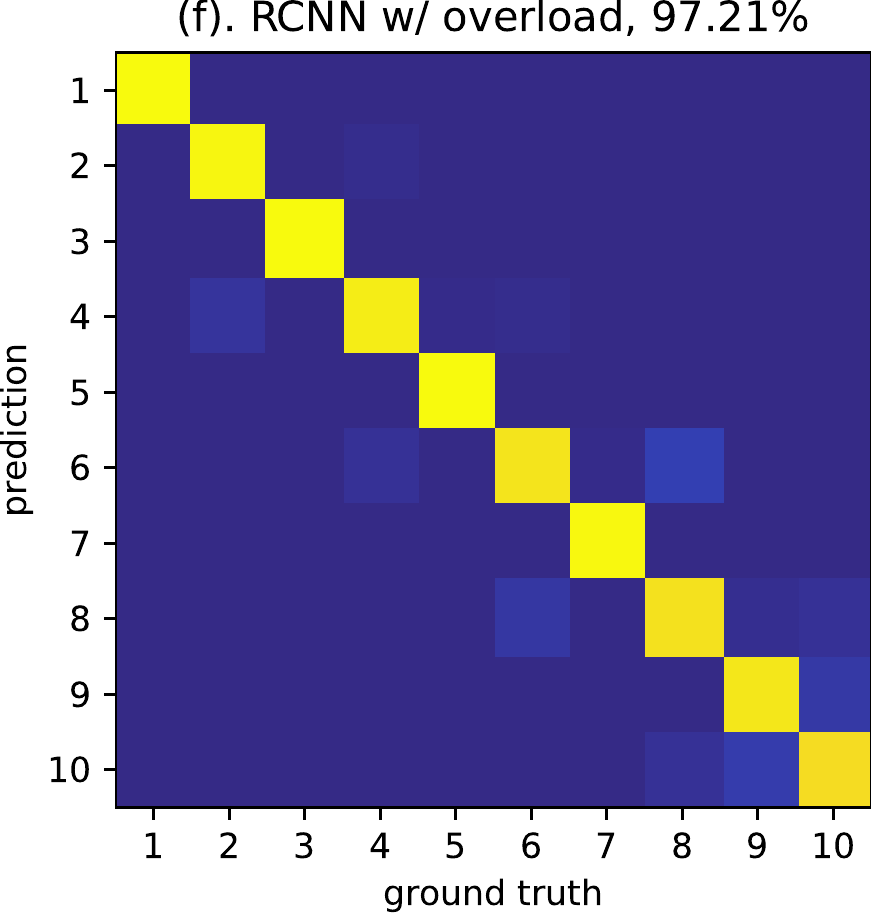}
 } 
 \subfigure{
 \includegraphics[width=0.22\linewidth]{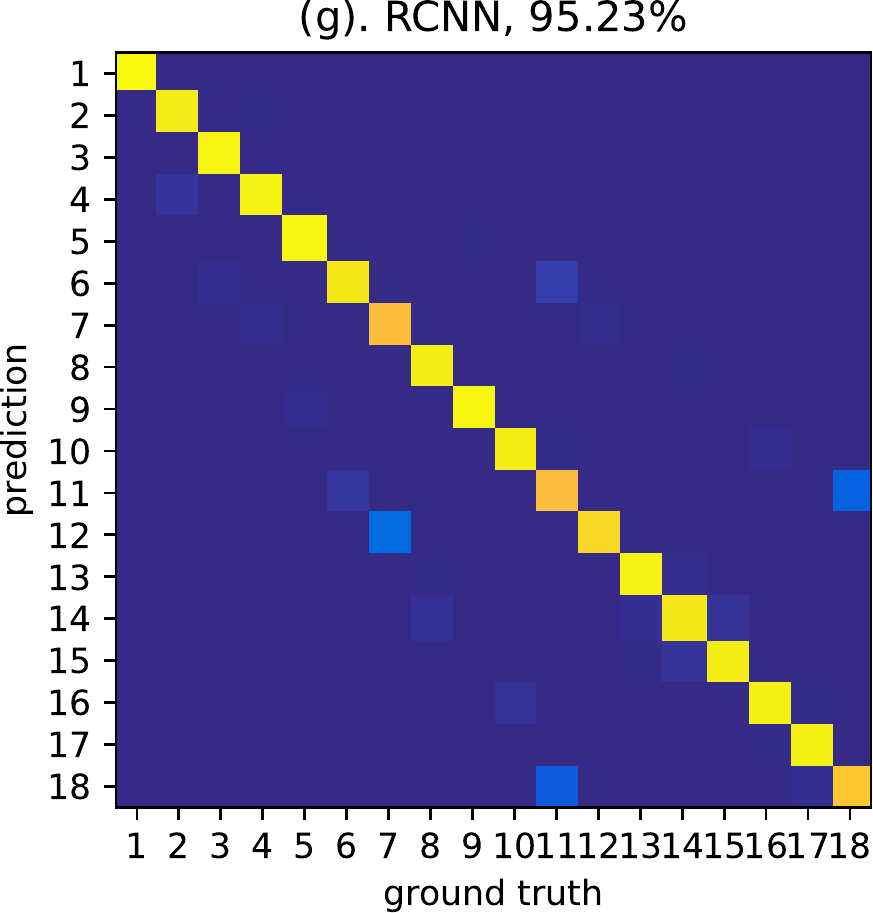}
 } 
  \subfigure{
 \includegraphics[width=0.22\linewidth]{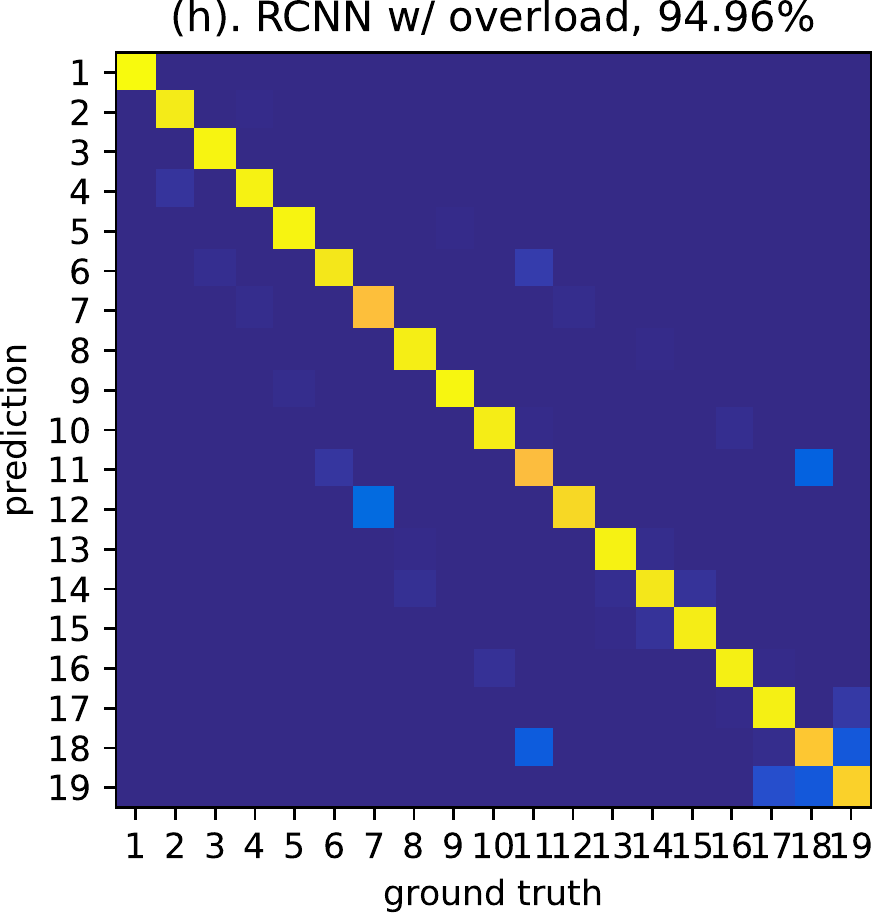}
 } 
 \caption{Confusion matrices for different classification tasks. Test data has an SNR varying over [-10,10]dB and uses the default cross-entropy loss function. Note that dark colors represent low values and light colors represent high values. SUb-figures (a)-(c) examine classification of the number of signals, (d) includes the overloaded case, (e)-(f) examine the Nine-class case with and without overloading respectively, while (g)-(h) represent the 18-class case with and without the overloaded case. }
\label{fig:tasks} 
\vspace{-0.2in}
\end{figure*}

\begin{figure}[t]
  \centering
  \includegraphics[width=0.40\textwidth]{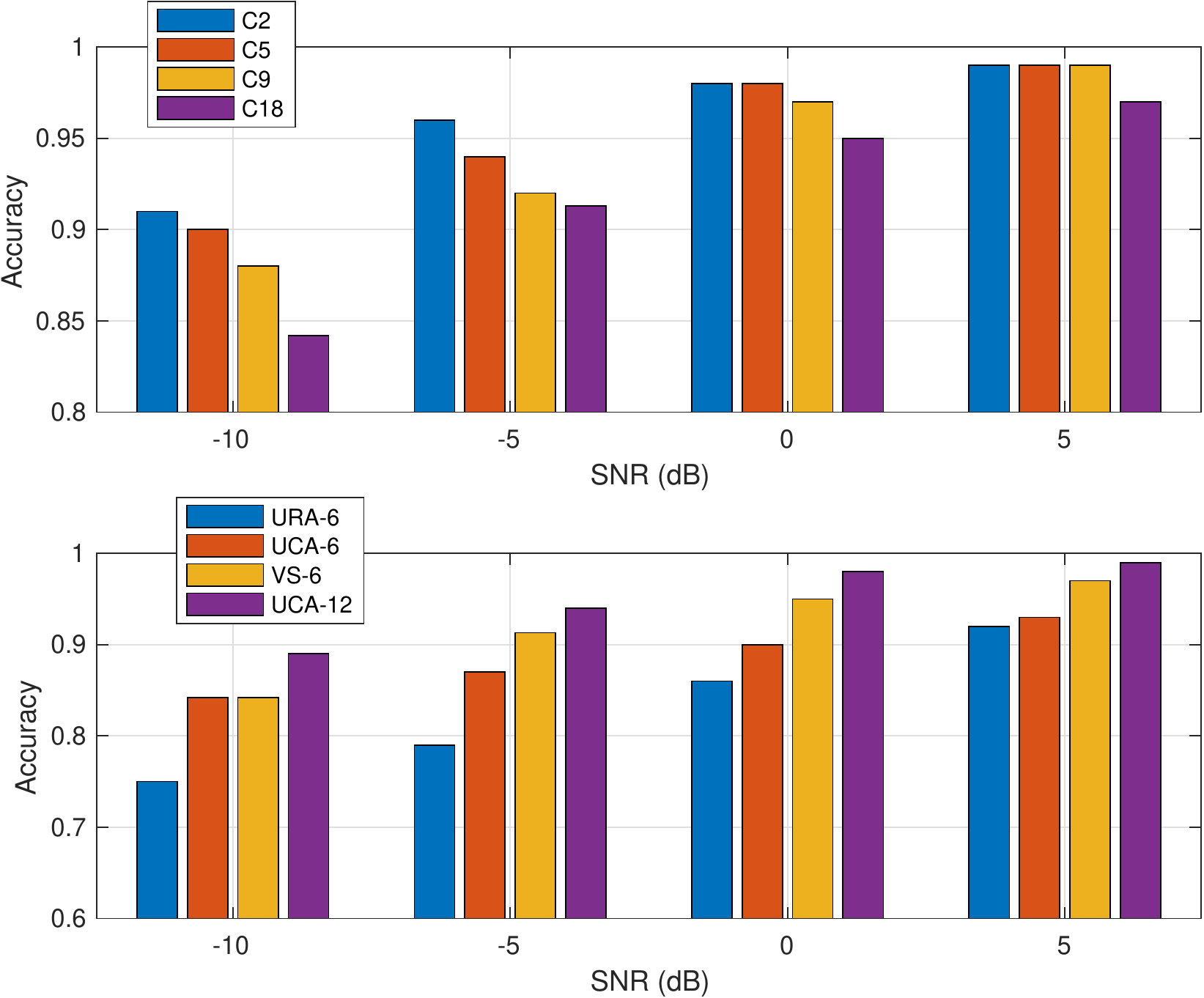}
  \caption{RCNN accuracy over SNR on different tasks and arrays}
  \label{fig:12_task_array}
\end{figure}

In our example scenario, we assume sources moving 
at speeds less than $100$\texttt{m/s} and a frequency of approximately $2.5$\texttt{GHz}, which corresponds to a channel coherence time $t_{coh}$ of $1.03$\texttt{ms}. 
For dataset generation, we assume a   channel with time-varying random phase due to propagation and movement.  We also assume perfect array geometry and neglect the effects of port offset and mutual coupling. 
Each entry in the dataset is $N=200$ samples long. 
We assume QPSK-modulated, fixed polarization pulse-shaped waveforms, sampled at 32 samples per pulse in order to simulate delays that can be either a large or small fraction of a symbol. 
It is assumed that the sources are uniformly distributed over a sphere, meaning that both azimuth and elevation angles are selected at random\footnote{Note that the elevation angles are selected such that the resulting points remain uniform over the sphere. If the elevation is simply selected in the interval $[-\pi, \pi]$, the points would appear to cluster at the poles. }.
We consider uniform random \emph{line of sight} SNR values in the interval $[-10,10]$\texttt{dB}. 
Multipath (i.e., non-line-of-sight) copies of signals are generated at power levels randomly between $[-3,0]$\texttt{dB} less than the line-of-sight component with a random delay uniformly distributed from 1 to 20 samples (i.e.,  coherent multipath interference), which are uniformly distributed over a cone centered on the LOS DoA with a span angle $30^\circ$. 
The model order is selected uniformly at random from the 18 options shown in Table \ref{table:MOE}. 
A comprehensive list of parameters is available in Table \ref{table:data}. 
We generate a dataset with $5\times 10^6$ labeled entries. 
The dataset is generated in Python. 
We use PyTorch to build the RCNN and SciPy to manipulate the network. 
All training and testing was conducted on a remote high-performance compute (HPC) system accelerated with GPUs. 
The largest model converged in 40 minutes on this system. 
Our data and code are available online\footnote{\texttt{https://github.com/yujianyuanhaha/MOE}}.

\begin{table}[tbpt]
\caption{Simulated Dataset Parameter Settings}
\label{table:data}
\begin{center}
\begin{tabular}{c|c}
\hline
 Setting & Range\\
\hline
 SNR & (-10,10)\texttt{dB}\\
\hline
 $N_s$ & 200\\
\hline
 coherent path signal power $p_i$ & $(-3,0)$\texttt{dB}\\
 \hline
 coherent path signal delay time $\tau_i$  & $(1,20) \frac{1}{f_s}$\\
\hline
 URA ($d_L$, $d_W$) & (0.1, 0.1) \\
\hline
 UCA $d$ & 0.2\\
\hline
azimuth $\theta$  &  $(0, 2\pi)$\\
\hline
elevation $\phi$  &  $(0, \pi)$\\
\hline
carrier frequency   &  $2.7$ \texttt{GHz}\\
\hline
modulation   &  QPSK\\
\hline
max coherence time   &  $1.03$ ms\\
\hline
generated datasize   &  $5 \times 10^6$\\
\hline
generated overloaded datasize   &  $3 \times 10^5$\\
\hline

\end{tabular}
\end{center}
\end{table}

\subsection{Performance Compared with Traditional Approaches}
We first compared the performance of the RCNN against the traditional estimators AIC and MDL. For this comparison, we examine the five-category performance since that is the design goal of the traditional approaches.
Fig. \ref{fig:tasks}(a)-(b) shows the confusion matrices for AIC and MDL while the RCNN is plotted in Fig. \ref{fig:tasks}(c).  Note that the RCNN Five-category labels are mapped from the 18-category prediction. Note that light colors represent high values (near $100\%$, while dark colors represent low values (near zero).  An ideal plot is dark everywhere except on the diagonal where it is yellow (i.e., an identity matrix).
It is apparent that the traditional approaches are not useful when coherent multipath is  present. 
Further, it can be seen in Fig.\ref{fig:tasks}(a)-(b) that distinguishing between four and five sources is particularly difficult for these techniques.  
In addition, both of these techniques tend to underestimate the model order rather than overestimate.  On the other hand, the RCNN is able to classify even the overloaded case as shown in \ref{fig:tasks}(d) where the overloaded case is specified as class 6.
Additionally, the RCNN approach is shown to provide high accuracy classification in all three classification tasks: 18-class (\ref{fig:tasks}(g)), Nine-class (\ref{fig:tasks}(e)), and Five-class. 
Note that all three tasks are handled by one model; the tasks are separated as in Table \ref{table:MOE} above. 
As each task is a refinement of the 18-class task, the accuracy improves with smaller numbers of classes. Like the Five-Class model, the Nine-Class and 18-class models are extended to include an overloaded category.  We find that the 18-class model, as well as the Nine-class mapping, can handle the overloaded case, something that the traditional models cannot do. We will discuss the overloaded case more shortly.

\subsection{Performance Compared with Other Deep Learning Models}

\begin{figure}[t]
  \centering
  \includegraphics[width=0.40\textwidth]{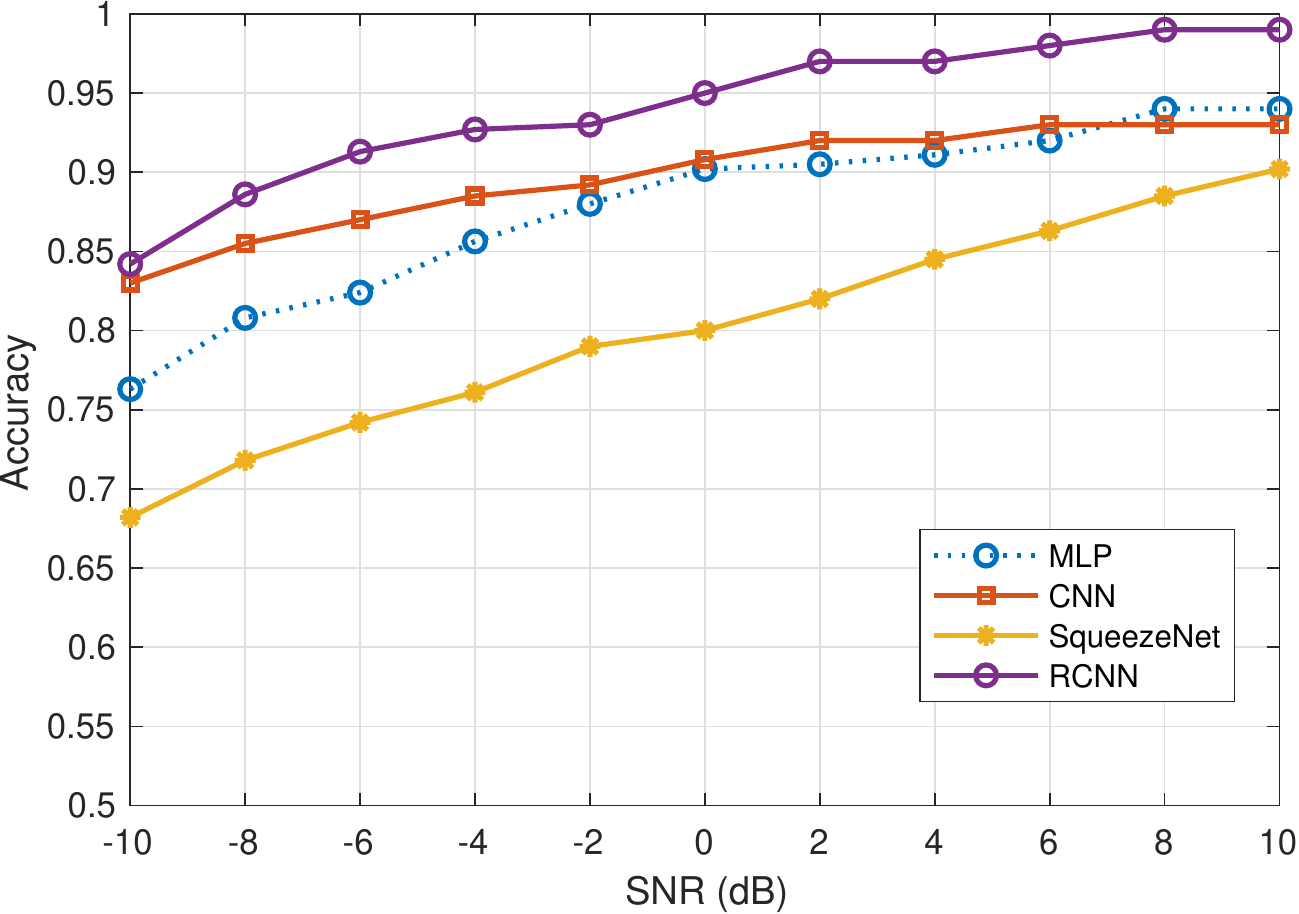}
  \caption{Different DNN model performance over SNR varying from $-10$\texttt{dB} to $10$\texttt{dB} for 18-category classification. Our proposed RCNN structure exhibits the best performance over this range. }
  \label{fig:11_DNN}
\end{figure}

Furthermore, we evaluate three different commonly seen deep neural network structures. 
\begin{itemize}
    \item MLP. MLP flattens data to one dimension, then traverses $[256,1024,512,256,128,64]$ fully connected layers each separated by batch normalization and dropout layers.  Such a MLP is adapted for MOE estimation in our earlier work \cite{yu2021direction}.
    \item CNN. We use six CNN layers with batch normalization, dropout, and maxpool layers between each. The output is flattened into one dimension.  Such CNN model is adapted in \cite{papageorgiou2020deep} for DoA estimation.
    \item SqueezeNet, which uses the same symmetric kernel grouped CNN but lacks residual links. We adapted open source code \cite{iandola2016squeezenet} \cite{squeezeNet} for this task. 
    \item RCNN, which is proposed above. 
\end{itemize}
Note that all these neural networks are trained and tested with 18-category labels, then we map the resulting labels into 9-categories (which indicates the number of signals and whether there is one source or multiple sources) or 5-categories (the number of signals) using the mapping function in Table. ~\ref{table:MOE}. We conclude from Fig. \ref{fig:11_DNN} that the RCNN exhibits the highest performance, followed by CNN, MLP, and SqueezeNet. 
Only RCNN and CNN show acceptable performance (deemed to be over $80\%$ at $-10$\texttt{dB}). We observe that the CNN has  better ability to understand the spatial correlation which exhibits matrix features, rather than the MLP which exploits vector features. RCNN performance is even better owing to the enhanced kernel representation using the asymmetric kernel, and learns  detail better due to the residual in the deeper layers.

\subsection{Performance with Different Array Geometries}

We evaluate our learning model on three common array geometries: the URA, UCA, and VS all with 6 elements
In addition, we also consider another more dense UCA with 12 elements. 
As discussed above we ignore the effects of mutual coupling and port offsets. 
Note that the 12-element UCA results in a $12\times 12$ complex covariance matrix, so the RCNN kernel sizes are changed to adapt to this experiment. 
From Fig. \ref{fig:12_task_array}, we can see that the VS exhibits the best performance of the 6 element arrays. 
As expected, the performance of the 12-element UCA exceeds the others, due to the greater number of degrees of freedom.

\subsection{Field of View}

The Field of View (FoV), or aperture, of an array describes the spatial degrees of freedom of the array. 
Commonly the DoA estimation variance is dependent on the DoA of the source. 
In other words, greater accuracy is possibly available at different DoAs. 
Here, we inspect the MOE classification performance across the FoV of a vector sensor array. 
In elevation, the error tends to increase near the poles and the performance is best near $90^\circ$. 
In azimuth, the error is minimized near $180^{\circ}$, and reaches a maximum near $0^{\circ}$. 
Overall, the error appears to be more dependent on elevation than azimuth. This can be explained by the fact that in VS manifold, azimuth and elevation play a different role.  For the same amount of slight change in azimuth and elevation, elevation  play a bigger role in real and imaginary value change in a covariance matrix.

\begin{figure}
  \subfigure{\includegraphics[scale=0.45]{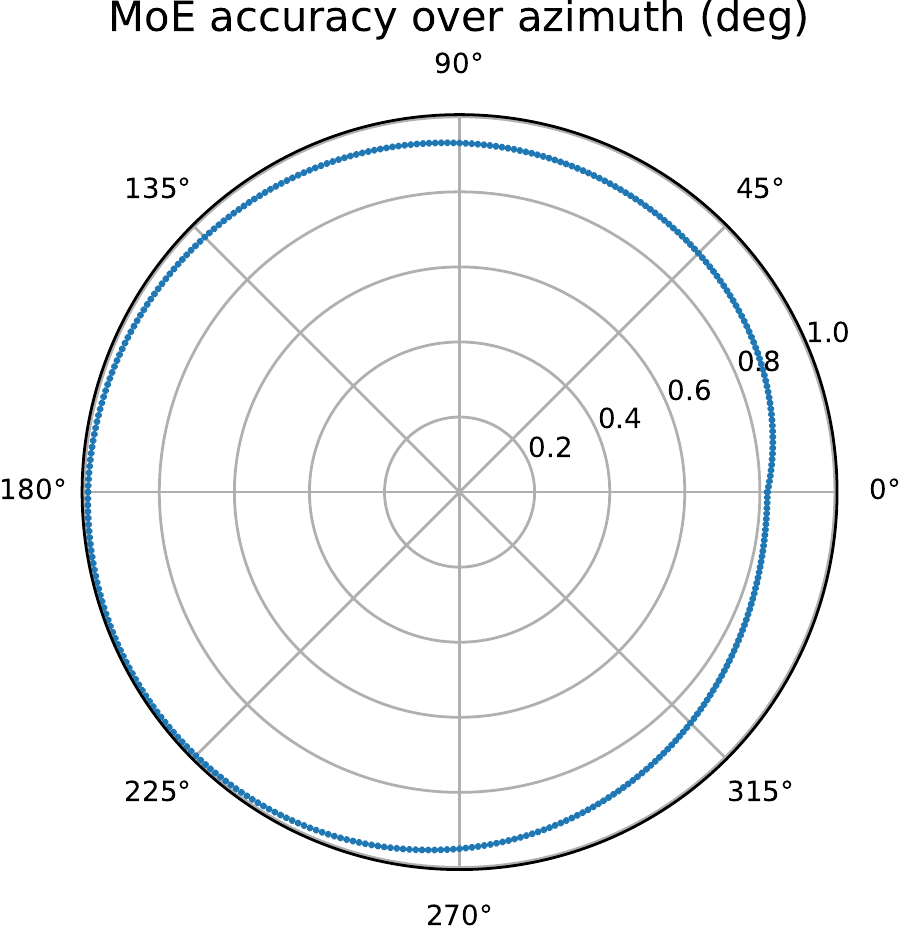}}
  \subfigure{\includegraphics[scale=0.45]{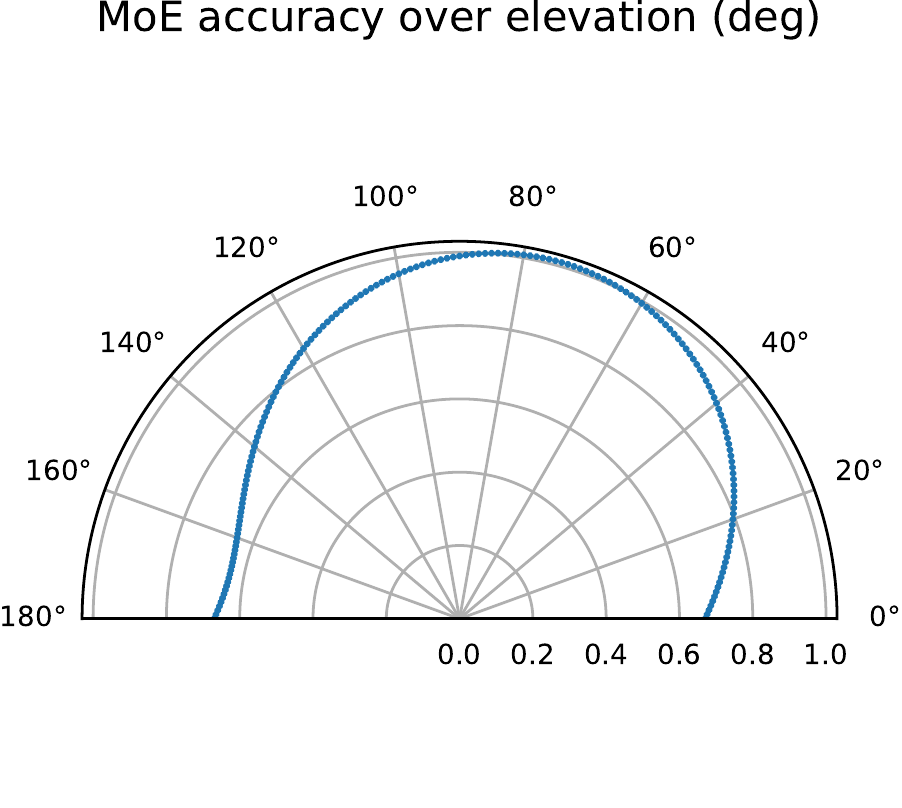}}
  \caption{MOE Accuracy of the proposed approach as a function of DoA azimuth angle (left) and elevation angle (right)}
  \label{fig:14_ele_azi}
\end{figure}

\subsection{Identifying Overloaded Scenarios}
In a more general case, the array can be \emph{overloaded}. 
This means that the true model order exceeds the theoretical limit, $n_M>5$ and $n_P>5$, for a 6-element array. 
We would like to have a classifier that can at least identify this scenario.  To do this, we can modify our classifier to include a 19$^{th}$ category. 
This category is labeled $19$ and corresponds to  cases when $n_M \in \{6, 7, 8\}$. 
The RCNN output layer dimension is changed from 18 to 19 to handle this additional class. 
Fig. \ref{fig:tasks} (h) shows that the new model suffers less than a $1\%$ performance penalty in the full classification problem. Similar results are obtained for the Five-class and Nine-class cases (Fig. \ref{fig:tasks} (d) and (f) respectively).
Further, the additional confusion is limited to uncertainty between the two largest classes for each task.

\subsection{Weighted Loss Function}

\begin{figure}
  \subfigure{\includegraphics[scale=0.45]{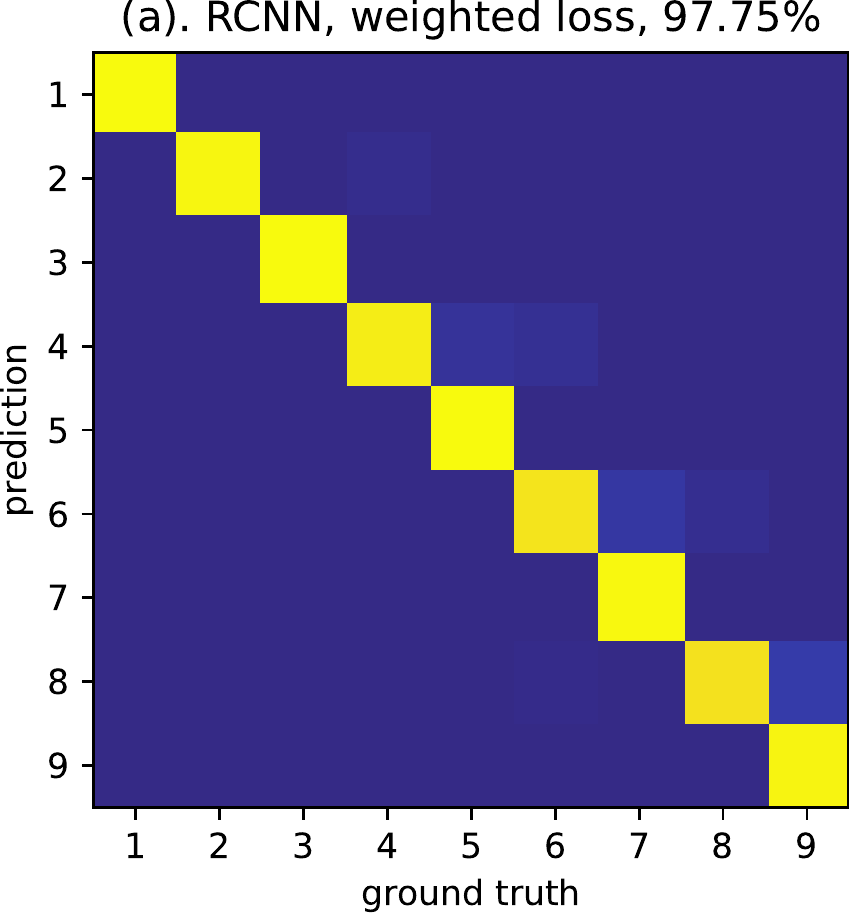}}
  \subfigure{\includegraphics[scale=0.45]{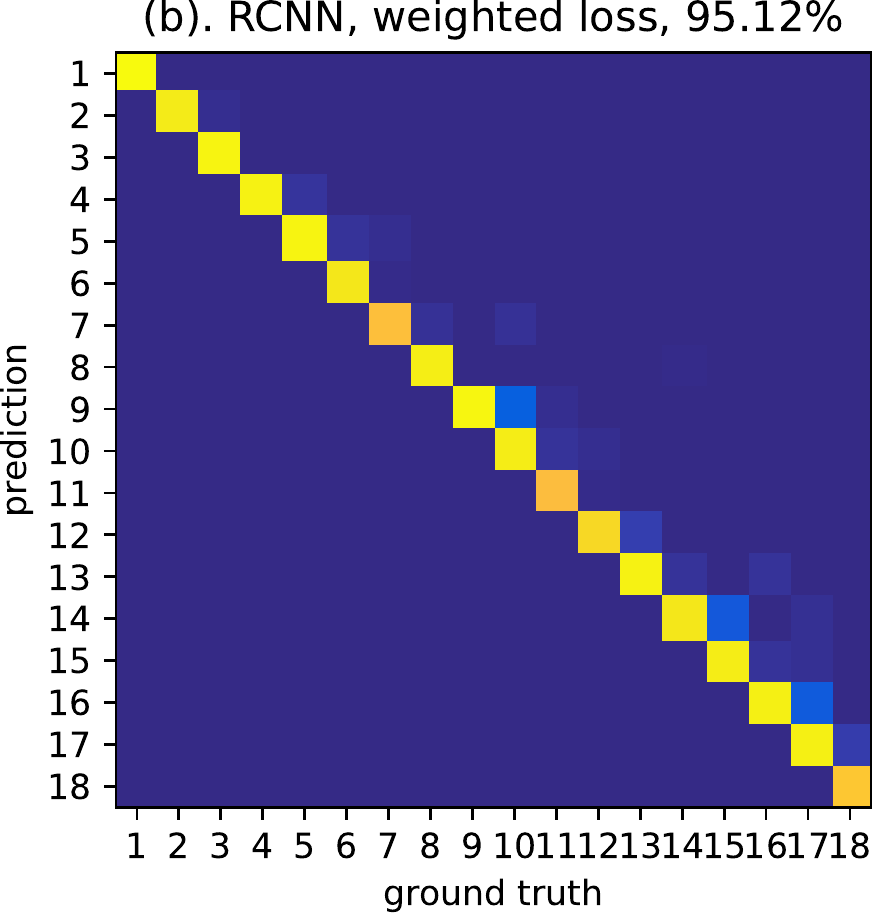}}
  \caption{RCNN with weighted loss function to eliminate the underestimation errors.}
  \label{fig:21_weight}
\end{figure}

When the weighted loss function is implemented in the RCNN, and no other changes are made, we observe that the network takes slightly more time to converge but converges to nearly the same accuracy. 
The confusion matrix shown in Fig. \ref{fig:21_weight} indicates that when errors occur, they are more often overestimation errors. 
In addition, the refined tasks (Five-class and Nine-class) suffer even less error than the 18-class task. 
Further, the network tends to overestimate the multipath label, suggesting the presence of multipath when there is none, and appears to avoid estimating no multipath when there is. 
These types of errors are more favorable than underestimation, since the subsequent estimators prefer overestimation.

\subsection{Computational Cost}

\begin{table}
\caption{Prediction accuracy \& Computation complexity} 
\centering 
\begin{tabular}{c | c  c  c } 
\hline
 Method & Accuracy (in \%)  &  params ($1e3$) & complexity ($1e6$ FLOPs)\\ 

\hline 
MLP        & 88.5 &  62.7 & 90.1 \\
CNN         & 92.8 &  117.9 & 120.8  \\
SqueezeNet & 84.1 &  10.9 & \textbf{ 26.5} \\
RCNN         & \textbf{95}.2 &  \textbf{60.9} & 42.2  \\

\hline 
\end{tabular}
\label{table:complexity}
\end{table}

\begin{figure}
  \centering
  \includegraphics[width=0.40\textwidth]{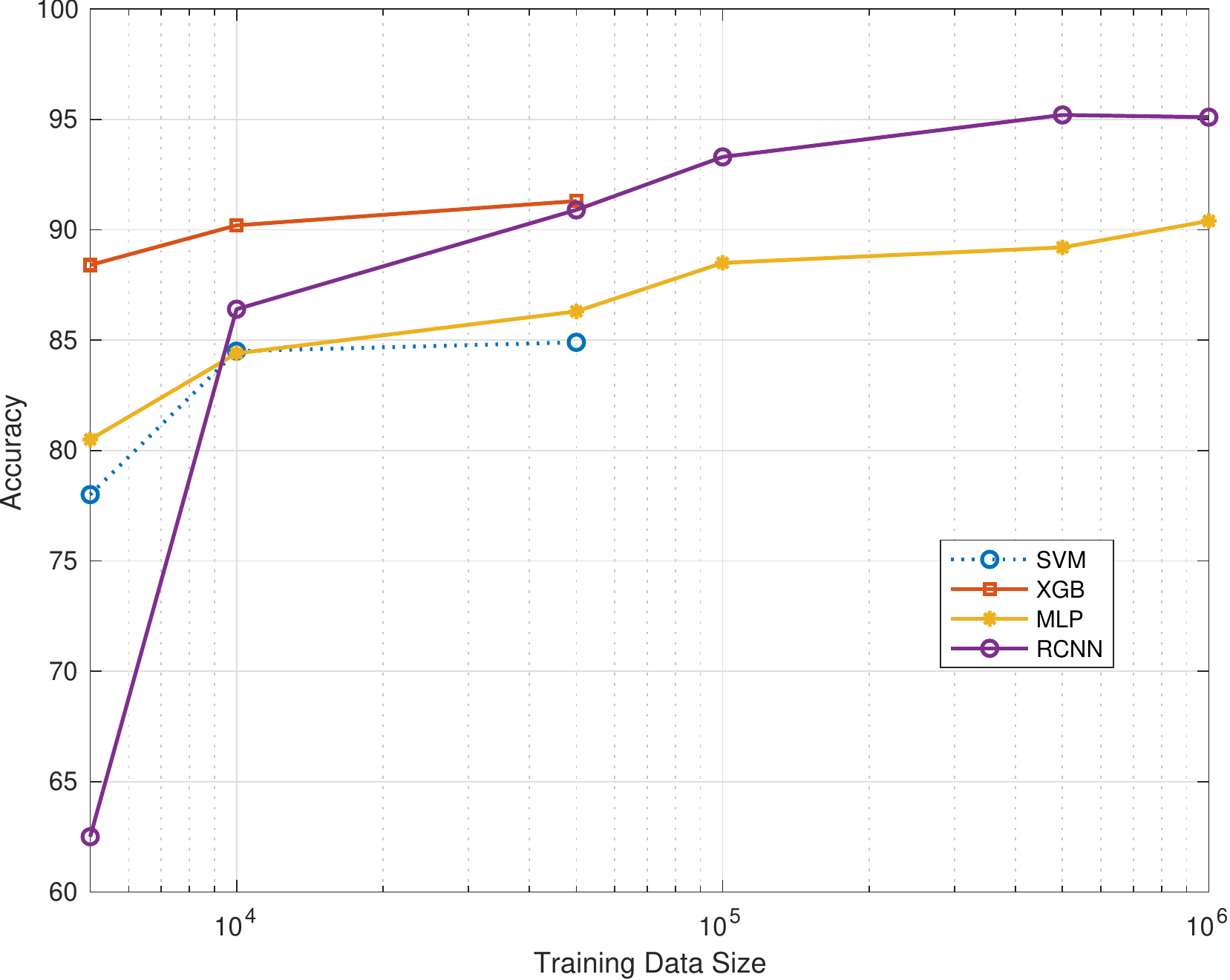}
  \caption{Performance vs. training data size.  Note that  SVM and XGB cannot be run with larger datasizes due to the large the memory requirements.   }
  \label{fig:15_trainSize}
\end{figure}

We demonstrate the relationship between training size and performance in Fig. \ref{fig:15_trainSize}. 
This compares the medium-sized networks SVM and XGM \cite{chen2015xgboost} as well as the heavier MLP,  and our proposed RCNN. 
With relatively medium-sized training data, we find that XGB exhibits the best performance. 
Once the data size exceeds $4\times 10^5$, note that XGB and SVM grow exponentially in time and memory, prohibiting them from further consideration. 
With large data sizes, we see that RCNN performance saturates near datasets with $4\times 10^5$ entries. 

Instead of deriving the theoretical computational cost of these techniques, we adapt the open-source tool flop-counter \cite{flops_counter} to estimate the complexity. 
We can see in Table \ref{table:complexity} that CNN has the highest computational cost, while RCNN balances complexity and performance the best. 
SqueezeNet with an asymmetric kernel has the lowest complexity but substantially reduced performance. However,  combining RCNN and SqueezeNet with  an asymmetric kernel can significantly reduce the computational cost compared to a single large kernel, and RCNN with a residual link  outperforms SqueezeNet significantly  with slightly improved computational cost.

\subsection{Benefit in DoA estimation}
\begin{figure}
  \centering
  \includegraphics[width=0.40\textwidth]{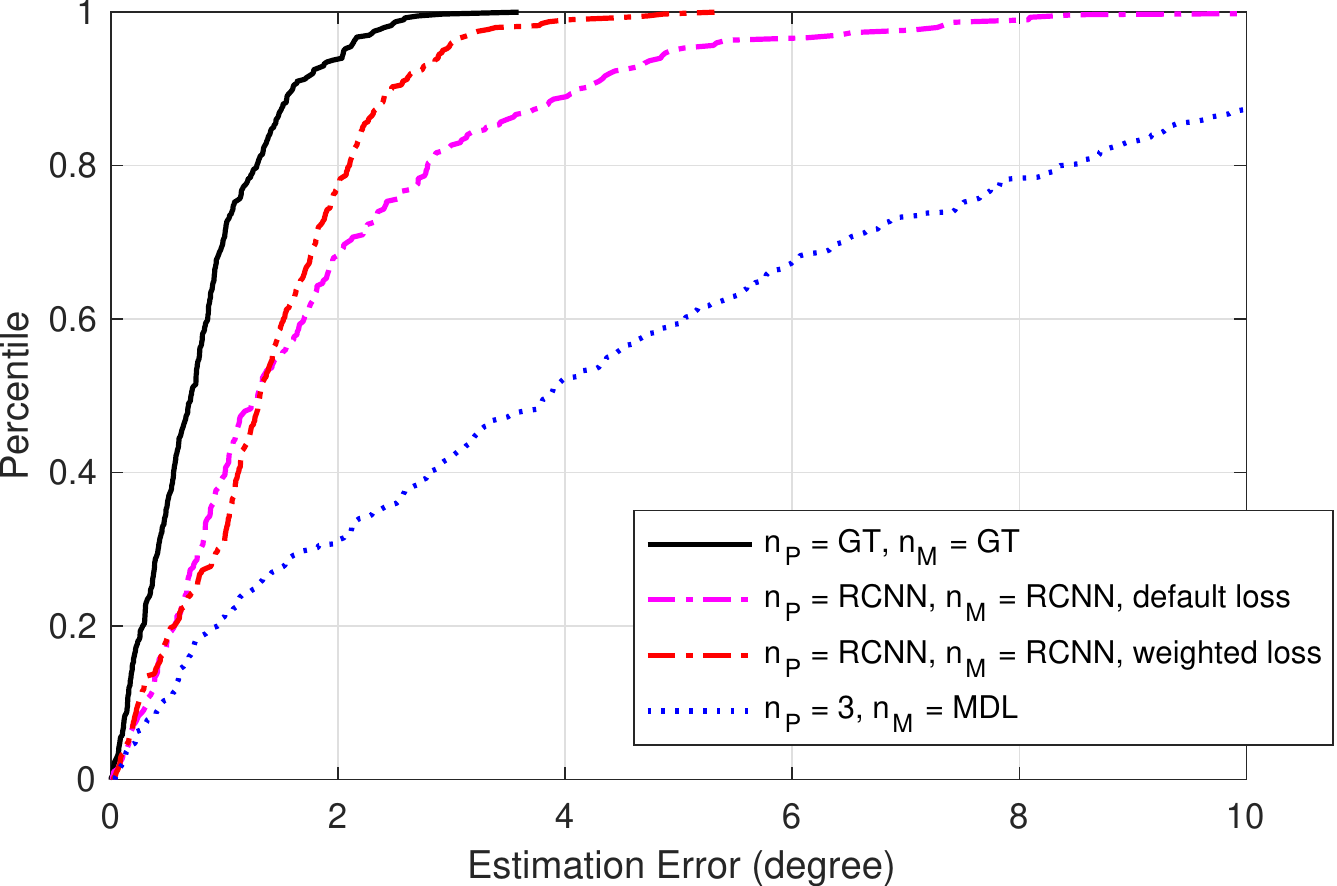}
  \caption{MUSIC with different MOE approaches, with $n_M \leq 3$, SNR $[-10,10]$dB }
  \label{fig:17_MUSIC}
\end{figure}

Eventually we study how path and model estimation from RCNN work along with MUSIC. As CDF plots shown Fig.~\ref{fig:17_MUSIC}, with test data from $n_M \in  [1,3]$ and $n_P \in [1,3]$, $SNR \in [0,10]dB$. We skip these higher numbers of modules here else  MUSIC estimation will take too much time. We have ground truth feeded MUSIC reach $1.6^{\circ}$ at 90th percentile as upper bound, while MDL preforming worse than 10 degree since it fails at MOE estimation. RCNN with weight loss can reach $2.4^{\circ}$ degrees at 90th percentile,  better than RCC with default loss. We can see the one with default loss results in many outliers, in that it has underestimation that largely harms  DoA estimation. On the other hand, the weight-loss RCNN work work slightly worse than ground-truth feeded, which show the DoA is fine with slight path or module overestimation.

\subsection{Reduced Computation in Signal Association}
Here we show how much gain our enhanced algorithm \ref{alg:sa} can achieve in signal association. With an assistant label, the number of path $n_P$ and the number of source $n_S$, we can largely reduce the number of signal correlation operations. For example,  12th category $(A,A,A,A,A)$ and 18th category $(A,B,C,D,E)$ already meet the goal of signal association, thus do not require any correlation operation. While for 13th case $(A,A,A,A,B)$ and 17th case $(A,A,B,C,D)$, they both require only at most 4 times correlation operation. In another way, using the default way requires $4+\dots+1 =10$ correlation operation, which largely slows down the efficiency of the tracking system. The computation is shown in Table \ref{table:sa}. For default algorithm only require $n_M$, the number of correlation operation is at most $\frac{n_M (n_M-1)}{2}$, while the algorithm whith $n_M$ and $n_S$ result in slightly smaller correlation count at most $\frac{n_S (n_S-1)}{2}$ since $n_S < n_M$. Meanwhile, our enhance signal association require the least correlation counts owing to the extra $n_P$ information that can narrow down the search space. In conclusion, we can see $n_S$ can reduce as much as 44\% off computation, while additional $n_P$ further reduce 13\% off computation.

\begin{table}
\caption{ Average Correlation Operation Counts in Signal Association} 
\centering 
\begin{tabular}{c |  c  c  c  c} 
\hline
Feed information & 2  &  3 & 4 & 5\\ 
\hline 
$(n_M)$        & 1.0 &  3.0 & 6.0 & 10.0 \\
$(n_S,  n_M)$        & 0.0 &  0.6 & 1.6 & 5.6 \\
$(n_S, n_M, n_P)$ & 0.0 &  0.6 &  1.6 & 4.3 \\
\hline 
\end{tabular}
\label{table:sa}
\end{table}

\section{Conclusions and Further Work}

In this work we have developed an RCNN model for Model Order Estimation for a passive array in the presence of coherent multipath.  We have also demonstrated how that model can be used in the context of a DoA system that determines the total number of signals present, the number of independent sources, and the association between signals and sources, even in the presence of coherent multipath.  The proposed  RCNN model provides better MOE estimation performance in the presence of coherent multipath interference then previously proposed approaches, both traditional and deep-learning-based.  The approach can provide the total number of paths incident on the array (correlated or uncorrelated), the number of sources, and the number of paths per source.  Further, we showed that a  weighted loss function can suppress the underestimation of the number signals/paths which provides a distinct advantage in DoA estimation. This model can give essential information for DoA estimation (using either MUSIC or other techniques) as well as the spatial filter of a localization/tracking system. 

A limitation of this work is the need for labeled training data.  Thus, future work includes examining transfer learning techniques such as an auto-encoder or optimal-transport (OT) approach that require less labeled training data, and are capable of adapting to new environments (e.g., minor antennas spacing offsets).




\bibliographystyle{IEEEtran.bst}
\bibliography{IEEEabrv,references.bib}

\end{document}